\documentclass[aps,prr,twocolumn,superscriptaddress,longbibliography]{revtex4-2}
\usepackage{amsmath}
\usepackage{amsfonts}
\usepackage{amssymb}
\usepackage{lmodern,dsfont}
\usepackage{graphicx}
\usepackage[usenames,dvipsnames]{xcolor}
\usepackage{bm}
\usepackage[english=american]{csquotes}
\usepackage[normalem]{ulem}
\usepackage{color}
\usepackage {cancel}

\renewcommand{\eqref}[1]{Eq.~(\ref{#1})}

\begin{document}

\author{Kenshin Matsumoto}
\affiliation{Department of Physics \#1, Graduate School of Science, Kyoto University, Kyoto 606-8502, Japan}
\author{Shin-ichi Sasa}
\affiliation{Department of Physics \#1, Graduate School of Science, Kyoto University, Kyoto 606-8502, Japan}
\title{Singularity of information flow at the Hopf bifurcation point}
\date{\today}

\begin{abstract}
We investigate the singular behavior of information flow near the Hopf bifurcation point by analyzing the learning rate as a specific measure of information flow. 
We study the Brusselator, a model system exhibiting the Hopf bifurcation.
We first numerically compute the learning rate in the stationary regime and find that it remains finite even in the deterministic limit, suggesting that the learning rate  can be quantified in deterministic dynamics through probabilistic descriptions. 
Linear analysis accurately reproduces the numerical results in the stationary regime but fails near the bifurcation point. 
To overcome this limitation, we employ the singular perturbation method, well known in deterministic bifurcation theory, and carry out the corresponding calculation explicitly for a stochastic system described by a Langevin equation. This allows us to evaluate the learning rate near the bifurcation point.
We then theoretically derive its non-smooth behavior in the deterministic limit. 
Our results demonstrate that changes in dynamical behavior are reflected in the learning rate  and provide a basis for analyzing information processing in biochemical oscillations.
\end{abstract}

\maketitle

\section{Introduction}
Stochastic thermodynamics \cite{Sekimoto2010} has evolved into information thermodynamics \cite{parrondo2015thermodynamics,Sagawa2019} integrating with information theory \cite{Shannon,coverthomas}.
The Szilard engine \cite{szilard} is known as the simplest model of Maxwell’s demon. This model has been used to quantify the work required for measurement by the demon.
Moreover, the minimal work needed to erase the information acquired through measurement is constrained by Landauer’s principle \cite{Landauer}.
More recently, general relations between work and information have been derived for feedback-controlled systems, based on the fundamental equivalence between information and energy \cite{Sagawa, sagawa_ueda_2010}.
In the past decade, various forms of the second law incorporating information have been proposed, leading to active discussions on the fundamental limits of information processing \cite{TUR,Tanogami_Saito_2023}.

One way of quantifying the information flows between two physical systems is the learning rate \cite{langevin_information,conti_information},
which is defined by decomposing the time derivative of the mutual information into contributions from each variable.
Because the mutual information quantifies correlations between variables, the learning rate represents the contribution of each variable to those correlations.
The learning rate appears in the second law of information thermodynamics for subsystems when the system satisfies the multipartite condition \cite{Multipartite_information_flow}.
This inequality implies that dissipation in the observed subsystem can be reduced by the information flow.
In particular, the information flow enables heat flow into the subsystem from the environment at steady state, or even the extraction of work from the subsystem.
Recently, learning rates have been used to formulate the effect of information flow and information-processing efficiency in noisy signal transduction systems \cite{Barato_2014,Hartich_2014,ito2015maxwell,Sensory_capacity,Role}, as well as in analyzing the impact of information flow on cooling and work extraction in heat engines \cite{Information_Arbitrage}.
Thus, the learning rate plays a key role in quantifying information flow across a wide range of systems, including biological and experimental setups.

Oscillatory phenomena emerge as a special class of dissipative structures in fluctuating systems \cite{Prigogine_Nicolis_1971}.
When a system exhibits oscillatory dynamics, the variables tend to follow each other in time, suggesting the presence of information flow between them.
Indeed, in intracellular chemical reaction networks such as gene regulatory systems, the presence of feedback—a fundamentally information-theoretic concept—can give rise to oscillations \cite{novak2008design}. Oscillations are ubiquitous in biological systems, such as in circadian rhythms \cite{cell_rythms, Circadian_Oscillation} and the cell cycle \cite{cell_cycle}.
Therefore, it is essential to clarify the relationship between oscillatory behavior and information flow for understanding biochemical reaction dynamics.

In biological systems, the dynamics of the molecular numbers in biochemical reactions are described by stochastic processes due to intrinsic fluctuations.
When the system size is small, the time evolution of the molecule numbers of chemical species is described by the master equation.
In the large system size regime, the concentration dynamics of chemical species are described by Langevin equations.
In particular, oscillatory phenomena in intracellular chemical reaction systems described by such stochastic processes are referred to as biochemical oscillations.
Extensive research has been conducted on the thermodynamic aspects of biochemical oscillations based on stochastic descriptions.
For example, in analyses based on the master equation, the transition of the thermodynamic quantities in systems exhibiting Hopf bifurcations \cite{phase_transition}, thermodynamic trade-off relations involving oscillation accuracy and robustness \cite{Coherence_of_biochemical, Design_priciples_for_biochemical,Robust_oscillations} have been studied.
Furthermore, in Langevin systems, the formulation of stochastic thermodynamics \cite{Stochastic_Thermodynamics_in_Mesoscopic}, the energetic cost required for accurate oscillations \cite{The_free-energy_cost, The_energy_cost_and_optimal_design} and the estimation of entropy production \cite{Improved_estimation_for_energy}  have been studied.
In this way, the thermodynamic understanding of biochemical oscillators described by stochastic processes has been steadily advanced.

In this study, we investigate the relationship between oscillatory behavior and information flow.
In particular, we focus on chemical reaction systems exhibiting a Hopf bifurcation, and analyze the singular behavior of the learning rate, which we use here as a specific measure of information flow, at the Hopf bifurcation point.
A Hopf bifurcation is a dynamical transition from a non-oscillatory state to an oscillatory one.
The Hopf bifurcation in deterministic dynamical systems is well understood within the framework of dynamical systems theory \cite{Strogatz,Kuramoto1984}.
For Hopf bifurcations in stochastic systems, various statistical quantities have been analyzed using descriptions based on phase and amplitude of oscillations \cite{Baras1982,Knobloch1983}, including stochastic averaging \cite{Arnold1996}.
In this paper, we employ the singular perturbation method to stochastic dynamics, in order to analyse information-theoretic quantities near the bifurcation point.
We demonstrate that the learning rate computed from numerical simulations is reproduced by the singular perturbation analysis. 
This confirms that the method is effective for analytically evaluating statistical quantities near the bifurcation point.
As our main result, we theoretically show a non-smooth change in the learning rate at the Hopf bifurcation point, which is not easily obtained by numerical simulations.

This paper is organized as follows.
Section II begins with the Langevin equation for the Brusselator, a chemical reaction system exhibiting a Hopf bifurcation, and its corresponding Fokker-Planck equation.
We provide an overview of the dynamical changes associated with the Hopf bifurcation.
We introduce the learning rate as a measure that quantifies information flow, and describe its role in the second law of information thermodynamics.
We then present numerical simulations of the learning rate in the steady state.
In Sec. III, we show that the results are accurately reproduced by the linear analysis in the non-oscillatory regime.
However, near the bifurcation point—including the oscillatory regime—linear analysis fails, and we demonstrate that the singular perturbation theory provides an accurate analytical description.
Based on the singular perturbation method for the Langevin equation, we show that the learning rate exhibits a singular change in the deterministic limit at the Hopf bifurcation point.
Section IV explains the details of the singular perturbation method applied to the Langevin equation.
Finally, Sec. V provides concluding remarks.

\section{Set-up}
\subsection{Brusselator}
As a representative chemical reaction system exhibiting a Hopf bifurcation, we introduce the Brusselator. The chemical reactions of the Brusselator are given by
\begin{equation}
\label{Brusselator reaction}
    \begin{matrix}
        A\rightarrow X_1,\\
        B+X_1\rightarrow X_2,\\
        2X_1+X_2 \rightarrow 3X_1,\\
        X_1 \rightarrow \phi,
    \end{matrix}
\end{equation}
where species $A$ and $B$ are supplied from particle reservoirs, and their concentrations $a$ and $b$ are assumed to be constant throughout the dynamics.
This Brusselator is known as an autocatalytic reaction system that exhibits chemical oscillations.
For the large system of volume $V$, the dynamics of the concentrations $\vec{x}=(x_1,x_2)$ of the chemical species $X_1$ and $X_2$ are described by the following Langevin equation \cite{Gillespie_langevin}:
\begin{equation}
\label{langevin}
        \dot{x}_i = F_i(\vec{x})+\sum^4_{j=1} B_{ij}(\vec{x})\cdot \xi_j(t),
\end{equation}
where $i = 1, 2$.
The drift term is given by
\begin{equation}
    \vec{F}(\vec{x})
    =\left(
    \begin{matrix}
        a-bx_1+x_1^2x_2-x_1\\
        bx_1-x_1^2x_2
    \end{matrix}
    \right),
\end{equation}
where all reaction rate constants are set to unity for simplicity.
The noise term $\vec{\xi}(t)$ represents a Gaussian white noise and satisfies the correlation property $\langle \xi_i(t) \xi_j(s) \rangle = \delta_{ij} \delta(t - s)$.
The matrix $B_{ij}(\vec{x})$ represents the noise intensity and is given by
\begin{equation}
    B(\vec{x})
    =
    \frac{1}{\sqrt{V}}
    \left(
    \begin{matrix}
        \sqrt{a}&-\sqrt{bx_1}&\sqrt{x_1^2x_2}&-\sqrt{x_1}\\
        0&\sqrt{bx_1}&-\sqrt{x_1^2x_2}&0
    \end{matrix}
    \right).
\end{equation}
Because $B(\vec{x})$ depends on the variables, the noise is generally multiplicative, and the symbol $\{\cdot\}$ in front of $\xi_j(t)$ in \eqref{langevin} represents the Ito product \cite{gardiner1985handbook}.
These dynamics can be described by the Fokker-Planck equation
for the probability density $p_t(\vec{x})$ and current $\vec{J}_t(\vec{x})$: 
\begin{equation}
\label{Fokker_Planck}
    \frac{\partial}{\partial t} p_t(\vec{x})
    =
    -
    \sum_{i=1}^{2} \frac{\partial}{\partial x_i} J_{t,i}(\vec{x}),
\end{equation}
where the probability current $\vec{J}_t(\vec{x})$ is given by
\begin{equation}
    J_{t,i}(\vec{x})
    =
    F_i(\vec{x})p_t(\vec{x}) - \frac{1}{2}\sum_j \frac{\partial}{\partial x_j} D_{ij}(\vec{x})p_t(\vec{x}).
\end{equation}
The diffusion matrix is defined as 
\begin{equation}
D_{ij}(\vec{x}) = \sum_{k=1}^4 B_{ik}(\vec{x}) B_{jk}(\vec{x}).
\end{equation} 
We treat $a$ as a fixed parameter and $b$ as a control parameter, following the usual convention for the Brusselator.

We briefly summarize the dynamical behavior of the Brusselator. 
In the deterministic limit $V \to \infty$, the dynamics of the Brusselator are described by the rate equation:
\begin{equation}
    \dot{x}_i
    =
    F_i(\vec{x}).
\end{equation}
The fixed point $\vec{x}_0$ of the deterministic system satisfies $\vec{F}(\vec{x}_0)=\vec{0}$.
For the Brusselator, the fixed point is given by $\vec{x}_0 = (a, b/a)$.
The Jacobian matrix of $\vec{F}(\vec{x})$ evaluated at the fixed point is defined as
\begin{equation}
\label{linear_operator}
    L_{ij} = \frac{\partial F_i}{\partial x_j}\Bigg\vert_{\vec{x}=\vec{x}_0}.
\end{equation}
The stability of the fixed point is determined by the eigenvalues of the Jacobian evaluated at that point.
When $b < a^2 + 1$, the real part of a complex-conjugate pair of eigenvalues of the Jacobian is negative, and the concentrations of each chemical species converge to the fixed point.
When $b > a^2 + 1$, the fixed point becomes unstable, and the concentrations depart from the fixed point and converge to a stable limit cycle.
The change in dynamics is known as a Hopf bifurcation.
The value of the control parameter at the bifurcation point is given by $b_c \equiv a^2 + 1$.

\subsection{Learning rate of Fokker-Planck equation}
In this study, we use the learning rate as a quantity that represents information flow.
We begin by defining the correlation between variables using the mutual information as
\begin{equation}
    I_t
    =
    \int d \vec{x}
    p_t(\vec{x})
    \ln \left(
    \frac{p_t(\vec{x})}{p^1_t(x_1)p^2_t(x_2)}
    \right),
\end{equation}
where $p^i_t(x_i)$ represents the marginal distribution of the variable $x_i$.
The mutual information has a non-negative value.
It becomes zero when the variables are statistically independent $p_t(\vec{x})=p^1_t(x_1)p^2_t(x_2)$.
Therefore, it characterizes the correlation between variables.
Using the Fokker–Planck equation \eqref{Fokker_Planck} and performing integration by parts under appropriate boundary conditions, we obtain the time derivative of the mutual information as
\begin{equation}
\begin{split}
    d_t I_t
    &=
    - \int d \vec{x}\vec{\nabla}^{\mathrm{T}}
    (\vec{\nu}_t(\vec{x})p_t(\vec{x}))
    \ln \left(
    \frac{p_t(\vec{x})}{p^1_t(x_1)p^2_t(x_2)}
    \right)\\
    &=
    \sum_{i=1}^2
    \left\langle
    \nu_{t,i} \partial_{x_i} \ln \left(  \frac{p_t}{p^i_t} \right)
    \right\rangle,
\end{split}
\end{equation}
where the local mean velocity $\nu_{t,i}$ is defined as 
\begin{equation}
\nu_{t,i} \equiv \frac{ J_{t,i}}{p_t(\vec{x})}.
\end{equation}
The information flow is defined by decomposing the time derivative of mutual information into contributions that correspond to each variable:
\begin{equation}
    l^i_t
    \equiv
    \left\langle
    \nu_{t,i} \partial_{x_i} \ln \left(  \frac{p_t}{p^i_t} \right)
    \right\rangle
    =
    \left\langle
    \nu_{t,i} \frac{\partial}{\partial x_i} \ln p^{|i}_t
    \right\rangle,
\end{equation}
where $p^{|i}_t$ denotes the conditional probability defined as $p^{|i}_t=p_t(\vec{x})/p^i_t(x_i)$.
This quantity is called the learning rate. It represents the contribution of each variable to the increase or decrease in correlation.
If $l^i_t > 0$, the change in $X_i$ leads to an increase in its correlation with the rest of the system.
This means that $X_i$ gains the information about other variables.
When the system satisfies the multipartite condition $(D_{ij} = 0\quad\mathrm{for}\quad \forall i \neq j)$, the learning rate appears in the second law of information thermodynamics \cite{conti_information}:
\begin{equation}
    \dot{S}^i_t+\dot{S}^{i,\mathrm{heat}}_t-l^i_t
    \geqq
    0,
\end{equation}
where $\dot{S}^i_t$ denotes the entropy change in the subsystem, and $\dot{S}^{i,\mathrm{heat}}_t$ denotes the entropy change due to heat flow.
When the system does not satisfy the multipartite condition, the learning rate appears in a more complicated form of the second law of information thermodynamics \cite{non-bipartite}:
\begin{equation}
    \dot{S}^i_t+\dot{S}^{i,\mathrm{heat}}_t-l^i_t
    \geqq
    \left\langle
    \frac{\nu_{t,i}}{D_{ii}}
    \sum_{j\neq i}D_{ij}\frac{\partial}{\partial x_j} \ln p_t
    \right\rangle.
\end{equation}
In this way, the learning rate serves as a quantitative measure of the information flow. 
In the present work, we mainly use the learning rate as an information-theoretic diagnostic of the temporal change of correlation and dynamical coupling in the nonequilibrium oscillator.
In this study, we analyze the learning rate of the system that undergoes the Hopf bifurcation, where the dynamics change from a non-oscillatory state to an oscillatory state at the bifurcation point.
\subsection{Numerical simulation}
\begin{figure}[tbp]
    \centering
    \includegraphics[width=80mm]{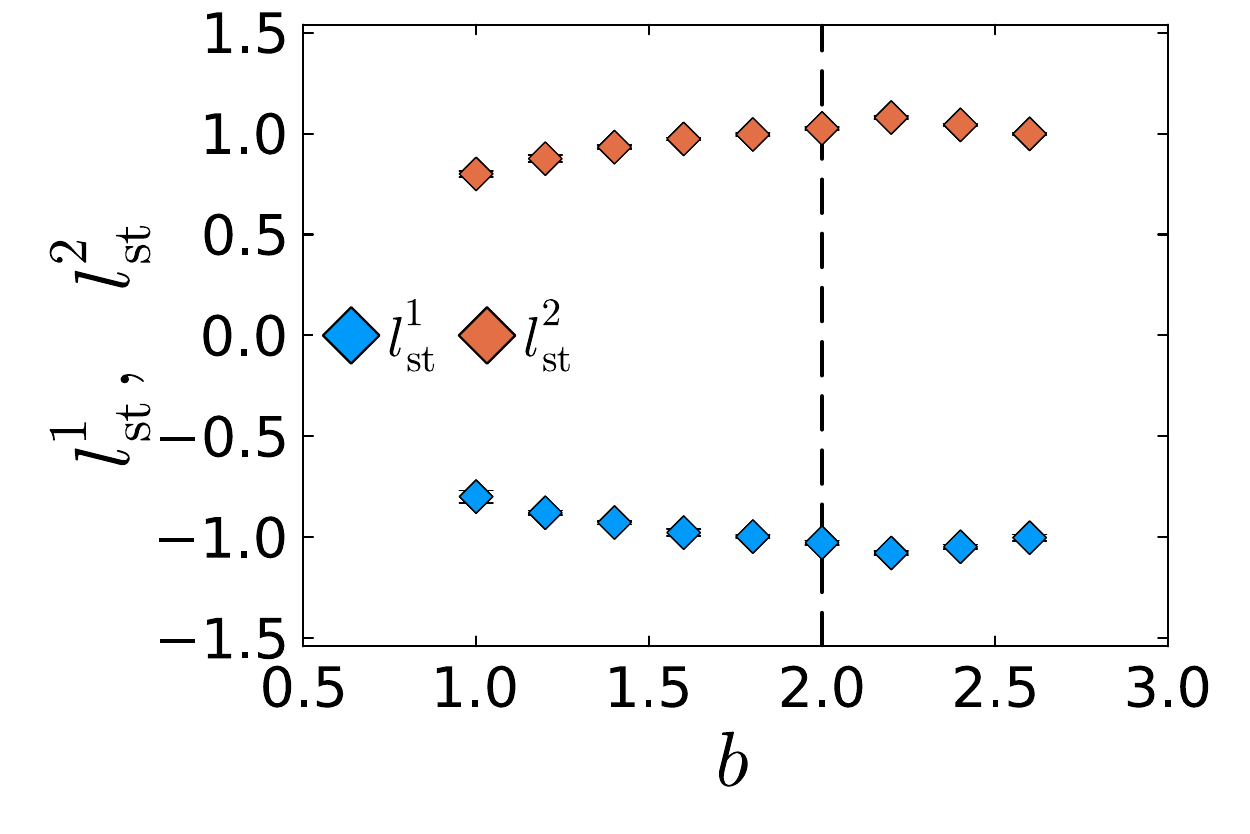}
    \caption{Learning rates in the steady state.
    The bifurcation point is \( b_c = 2 \), and the system size is \( V = 10^3 \).}
    \label{fig:simulation}
\end{figure}

We first perform numerical simulations of the Brusselator using the Euler–Maruyama method. Based on these simulations, we compute the steady-state learning rates $l^{1}_{\mathrm{st}}$ and $l^{2}_{\mathrm{st}}$.
The results are shown in Fig.~\ref{fig:simulation}.
First, because we focus on the steady state, we see the following relation:
\begin{equation}
    d_t I(X_1:X_2)
    =
    l^{1}_{\mathrm{st}} + l^{2}_{\mathrm{st}}
    =0.
\end{equation}
In the stationary regime $b<b_c$, the absolute value of the learning rates increases as the system approaches the bifurcation point.
Far from the bifurcation point on the oscillatory side $b>b_c$, the absolute value of the learning rate decreases.
This means that in both the stationary and oscillatory regimes, the learning rate can take the same value.
Therefore, it is impossible to determine whether the system is in the stationary or oscillatory regime based solely on the value of the learning rate.
Moreover, the singular behavior of the learning rate at the bifurcation point is not clearly observed in the numerical simulation.
In the following, we conduct a theoretical analysis to determine whether a singularity appears or not. 
\section{Results}
\subsection{Linear analysis in the stationary region}
As the first step, we perform a linear analysis in the stationary regime as a straightforward approach to reproduce the numerical results of the learning rate.
We introduce a new variable $\vec{\eta}$ as 
\begin{equation}
\label{intro_eta}
    \vec{\eta} \equiv \sqrt{V}\left(\vec{x}-\langle \vec{x}\rangle \right),
\end{equation}
where $\langle \cdot \rangle$ denotes the expectation with respect to the probability distribution $p_t(\vec{x})$, and $\langle \vec{x} \rangle$ represents the mean value of $\vec{x}$. Let $\hat{p}_t(\vec{\eta})$ denote the probability distribution of $\vec{\eta}$. Then, from \eqref{intro_eta}, we have
\begin{equation}
\label{intro_eta_pro}
\hat{p}_t(\vec{\eta}) = V p_t\left(\langle \vec{x} \rangle + \frac{1}{\sqrt{V}} \vec{\eta} \right).
\end{equation}
By substituting \eqref{intro_eta} and \eqref{intro_eta_pro} into \eqref{Fokker_Planck} and expanding in terms of the large system size $V$, we obtain the Fokker–Planck equation for $\hat{p}_t(\vec{\eta})$ to the lowest order:
\begin{equation}
\label{linear_Fokker}
\partial_t \hat{p}_t(\vec{\eta}) = - \sum_{i=1}^2 \sum_{j=1}^2 \left( \hat{K}_{t,ij} \eta_j - \frac{1}{2}\hat{D}_{t,ij} \partial_{\eta_j}  \right) \hat{p}_t(\vec{\eta}),
\end{equation}
where the drift matrix 
\begin{equation}
    \hat{K}_{t,ij}
    =
    \frac{\partial F_i}{\partial x_j}\Bigg\vert_{\vec{x}=\langle \vec{x} \rangle}
\end{equation}
and the diffusion matrix 
\begin{equation}
    \hat{D}_{t,ij}
    =
    D_{t,ij} (\langle \vec{x} \rangle)
\end{equation}
are independent of the variable $\vec{\eta}$. The resulting equation is called a linear Fokker–Planck equation, and when the initial distribution is Gaussian, the probability distribution remains Gaussian at all times. 
Consequently, statistical quantities can be expressed analytically.
In the case of the Brusselator, the learning rate $\hat{l}^{1}_{\mathrm{st}}$ for $\eta_1$ in the steady state is given by
\begin{equation}
\label{Brusselator_linear_learning}
    \hat{l}^{1}_{\mathrm{st}}
    =
    \frac{-4a^2b}{(b+b_c)^2-4a^2b}
\end{equation}
(see Appendix \ref{appendix_linear_learning}). 
Moreover, due to the property of the learning rate, it is not affected by parallel translations or constant rescalings of variables.
Therefore, the learning rate for the original variable $x_1$ coincides with that for the variable $\eta_1$.
\begin{figure}[tbp]
    \centering
    \includegraphics[width=80mm]{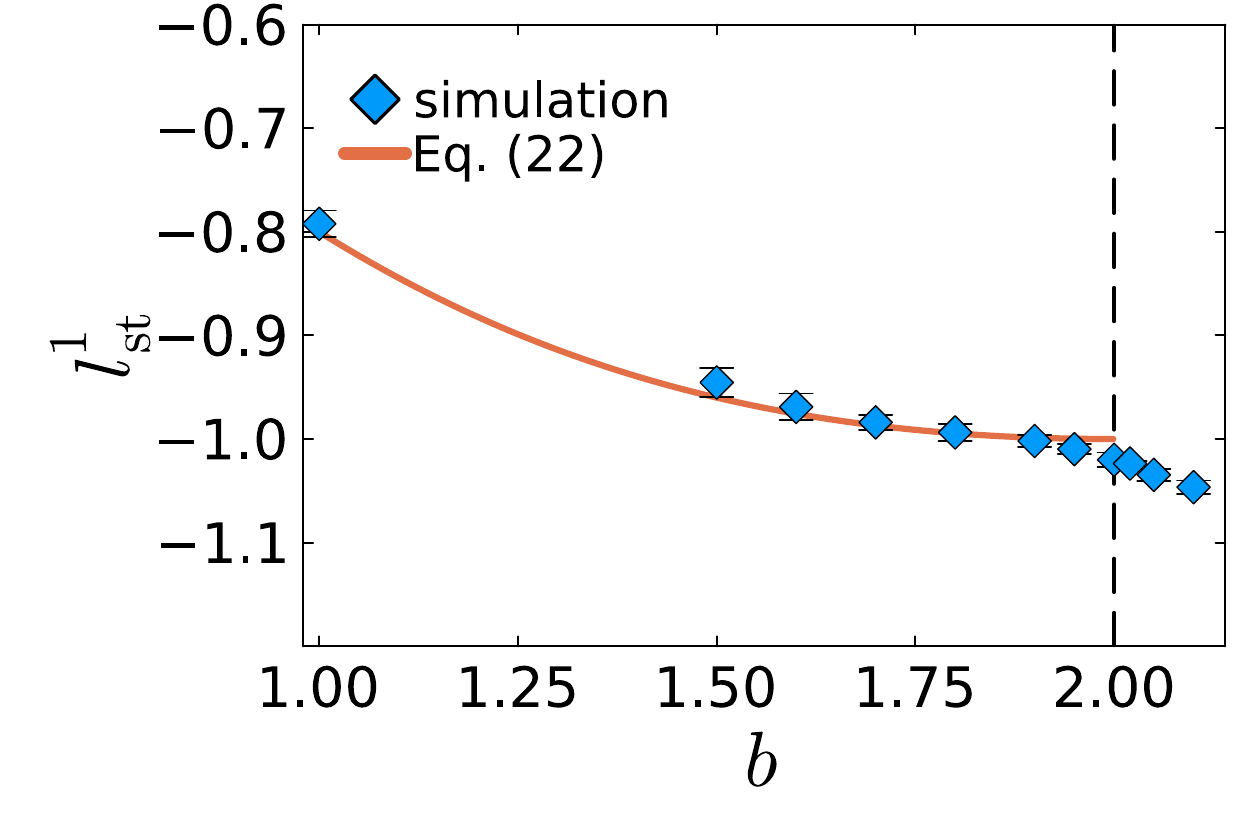}
    \caption{Learning rate in the steady-state regime. $b_c = 2$ and $V = 2 \times10^3$. Blue dots indicate the results of numerical simulations, and the orange line shows the result of the linear analysis given by \eqref{Brusselator_linear_learning}.}
    \label{fig:simulation_linear_analysis}
\end{figure}
We compare the learning rate given in \eqref{Brusselator_linear_learning} with the numerical simulation results in Fig. \ref{fig:simulation_linear_analysis}.
In the stationary regime, the learning rate calculated by the linear analysis accurately reproduces the results obtained by numerical simulations.
Because the linear analysis around the fixed point is valid only in the stationary regime, the corresponding curve is shown only up to the bifurcation point.
As $b$ approaches the bifurcation point, a deviation begins to emerge between the learning rate obtained by the linear analysis and that from the numerical simulations.
In the vicinity of the bifurcation point, the system exhibits qualitatively different behavior that is not captured by the linear analysis.
Therefore, we extend the deterministic bifurcation analysis to the stochastic dynamics described by the Langevin equation and analytically investigate the learning rate near the bifurcation point.

An important property of the learning rate should also be noted. The linear analysis is valid in the stationary regime when the system size is large.
The learning rate obtained in this regime remains nonzero even in the limit $V\rightarrow \infty$.
In other words, while information flows cannot be directly defined in deterministic systems, redefining the system as a stochastic process enables the definition of the learning rate.
A nonzero value of the learning rate in the deterministic limit indicates that correlations among fluctuations around the stable fixed point continue to be created and destroyed at a finite rate, even as the amplitude of the fluctuations vanishes.
This provides a clue for analyzing the dynamics from the perspective of information flow.

\subsection{Singular perturbation method for Langevin equation}
Near the bifurcation point of a deterministic equation exhibiting the Hopf bifurcation, the bifurcation analysis method has been established \cite{Kuramoto1984}.
In this theoretical method, a small parameter is introduced with respect to the bifurcation parameter, and by applying the singular perturbation, one can simultaneously derive the time evolution equation in which the limit cycle becomes circular.
This equation is called the normal form near the Hopf bifurcation.
One also obtains the coordinate transformation from the normal form coordinate to the original coordinates.

We introduce a bifurcation parameter $\mu$ to quantify the distance from the critical value of the control parameter $b$.
We assume that the real part of the eigenvalues of the Jacobian increases monotonically with respect to the bifurcation parameter $\mu$.
In the Brusselator, the bifurcation parameter $\mu$ is defined in terms of the control parameter just above the bifurcation point as
\begin{equation}
\label{bifurcation_parameter}
    \mu \equiv \frac{b-b_c}{b_c}.
\end{equation}
The small non-negative parameter $\epsilon$ is introduced in relation to the bifurcation parameter $\mu$ as
\begin{equation}
    \mu=\epsilon^2 \chi,
\end{equation}
where $\chi = \mathrm{sgn}(\mu)$ denotes the sign of the bifurcation parameter.
In this paper, we apply a perturbation theory to the Langevin equation by expanding with the small parameter $\epsilon$.
\begin{figure}[htbp]
    \centering
    \includegraphics[width=80mm]{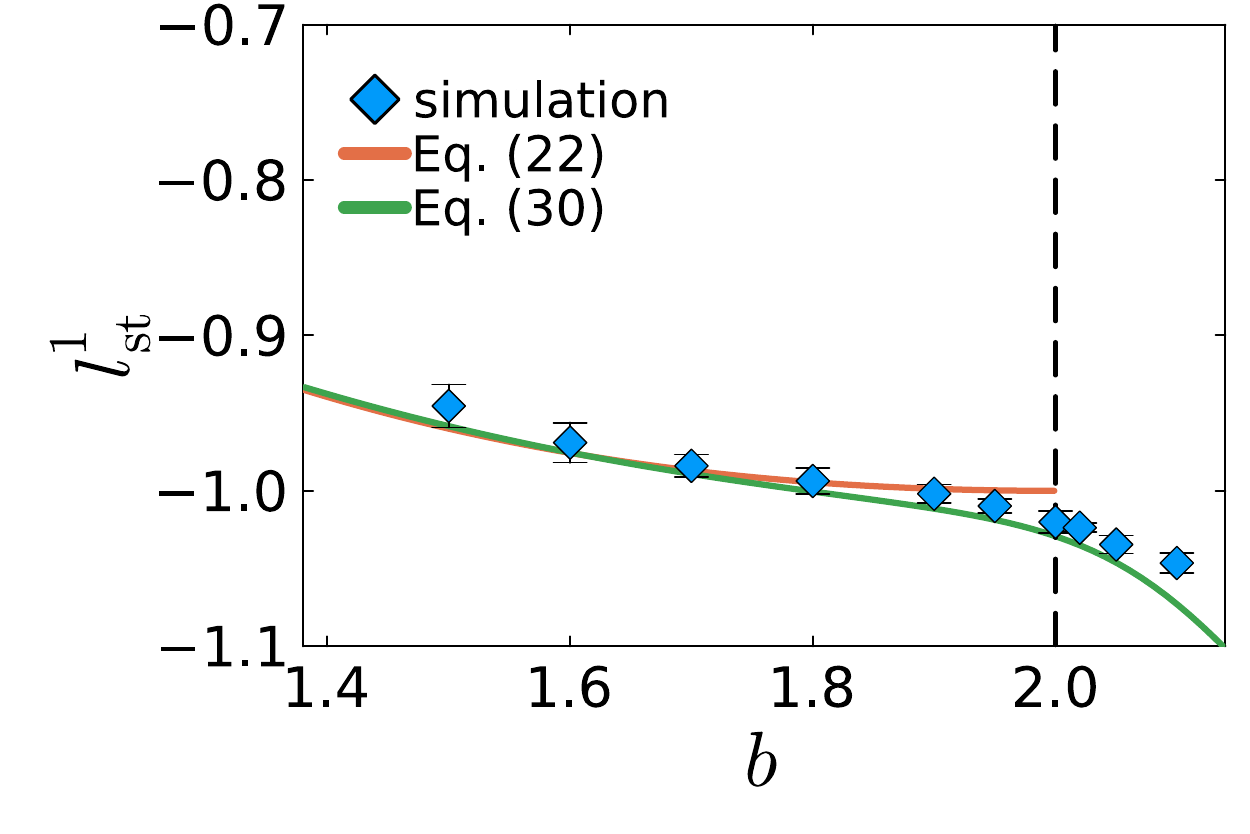}
\caption{Learning rate near the Hopf bifurcation in the Brusselator. $b_c = 2$ and $V=2\times10^3$. Blue dots represent the numerical simulation results, the orange line denotes the linear analysis result \eqref{Brusselator_linear_learning}, and the green line shows the result from the singular perturbation method \eqref{transformation_learning}.}
\label{fig:singular_perturbation}
\end{figure}

For stochastic Hopf bifurcations, related analyses have often been developed in amplitude--phase variables \cite{Baras1982,Knobloch1983}.
A commonly used example is stochastic averaging, where the phase dependence is reduced to simplify the stationary statistics \cite{Arnold1996}. 
To analyze the learning rate, we apply the singular perturbation to the Langevin equation and explicitly compute both the stochastic normal form that inherits the noise structure of the original Langevin equation and the higher-order terms of the coordinate transformation to the normal form variables.
To distinguish the normal coordinates from the original coordinates $\vec{x}$, we introduce the normal coordinates in boldface as $\boldsymbol{y}=(y_1,y_2)$.
As a result of the singular perturbation method, we obtain the stochastic equation for the normal coordinate $\boldsymbol{y}$ and the corresponding Fokker-Planck equation for the probability density of the normal form coordinate $\tilde{p}(\boldsymbol{y})$ written as
\begin{equation}
\label{normal_Fokker}
\begin{split}
    \frac{\partial}{\partial t}\tilde{p}(\boldsymbol{y})
    &=
    -\frac{\partial}{\partial \boldsymbol{y}}\cdot\left(
    \tilde{\boldsymbol{F}}(\boldsymbol{y})-\frac{1}{2}\frac{1}{\epsilon^2V} \tilde{D} \frac{\partial}{\partial \boldsymbol{y}}
    \right)\tilde{p}(\boldsymbol{y})\\
    &=-\frac{\partial}{\partial \boldsymbol{y}}\cdot(\tilde{\boldsymbol{\nu}}_t(\boldsymbol{y})\tilde{p}_t(\boldsymbol{y})),
\end{split}
\end{equation}
where we define the local mean velocity in the normal form coordinate as
\begin{equation}
    \tilde{\boldsymbol{\nu}}_t(\boldsymbol{y})
    \equiv
    \frac{1}{\tilde{p}(\boldsymbol{y})}
    \left(
    \tilde{\boldsymbol{F}}(\boldsymbol{y})-\frac{1}{2}\frac{1}{\epsilon^2V} \tilde{D} \frac{\partial}{\partial \boldsymbol{y}}
    \right)\tilde{p}(\boldsymbol{y}).
\end{equation}

As shown in Sec. IV, both the drift vector and the diffusion matrix are expressed in terms of the original system parameters through the singular perturbation analysis.
The stationary distribution of the Fokker–Planck equation \eqref{normal_Fokker} is derived by first considering the asymptotic region $V \gg 1$, and then performing a perturbative expansion in $\epsilon$. It is expressed by
\begin{widetext}
\begin{equation}
\label{stationary_distribution}
\begin{split}
    \tilde{p}_{\mathrm{st}}(y_1,y_2)
    &=
    C_0
    \exp \Bigg\{
    \epsilon^4 V \left[
    \frac{2 \chi \sigma_1}{\tilde{D}_{11} + \tilde{D}_{22}} (y_1^2 + y_2^2)
    - \frac{g_1}{\tilde{D}_{11} + \tilde{D}_{22}} (y_1^2 + y_2^2)^2
    \right]
    \Bigg\},
\end{split}
\end{equation}
\end{widetext}
where $C_0$ is the normalization constant and, $\omega_0,\sigma_1$ and $g_1$ are constants determined by the model (see Appendix \ref{stationary}).

We introduce a new variable 
\begin{equation}
\label{intro_u}
    \vec{u} \equiv \vec{x} - \vec{x}_0,
\end{equation} 
which shifts the fixed point to the origin. Because the transformation preserves the shape of the distribution, $\vec{u}$ and the original variable $\vec{x}$ share the same statistical properties.
The transformation from the normal form coordinate $(y_1, y_2)$ to the original coordinates is given by 
\begin{equation}
\label{intro_transfomration}
    \vec{u} = \vec{f}(\boldsymbol{y}),
\end{equation}
where $\vec{f}(\boldsymbol{y})$ is a polynomial function of $\boldsymbol{y}$.
In the normal form coordinate, the symmetry of the system allows the stationary distribution $\tilde{p}_{\mathrm{st}}(\boldsymbol{y})$ to be obtained analytically.
Under a general coordinate transformation, the learning rate can be calculated as 
\begin{widetext}
\begin{equation}
\begin{split}
\label{transformation_learning}
    l^{i}_{\mathrm{st}}
    &=
    \int d\boldsymbol{y}
    \sum_{kl}
    R_{ik}(\boldsymbol{y})
    \tilde{\nu}_{\mathrm{st},k}(\boldsymbol{y})
    \left[
    \frac{\partial}{\partial y_l}
    \ln{\frac{\tilde{p}_{\mathrm{st}}(\boldsymbol{y})}{\vert\det R(\boldsymbol{y})\vert}}
    \right]
    R^{-1}_{li}(\boldsymbol{y})
    \tilde{p}_{\mathrm{st}}(\boldsymbol{y}),
\end{split}
\end{equation}
\end{widetext}
where $R$ is the Jacobian matrix of the coordinate transformation $\vec{u} = \vec{f}(\boldsymbol{y})$ with components $R_{ij} \equiv \partial f_i(\boldsymbol{y}) / \partial y_j$ (see Appendix \ref{appendix_learning_transformation}).
Here, $l^{i}_{\mathrm{st}}$ depends on $\epsilon$ through $R(\boldsymbol{y})$, $R^{-1}(\boldsymbol{y})$, $\tilde{p}_{\mathrm{st}}(\boldsymbol{y})$ and $\boldsymbol{\tilde{\nu}}_{\mathrm{st}}(\boldsymbol{y})$, and it depends on $V$ through $\tilde{p}_{\mathrm{st}}(\boldsymbol{y})$ and $\boldsymbol{\tilde{\nu}}_{\mathrm{st}}(\boldsymbol{y})$.
We compare the results obtained from the singular perturbation theory, the linear analysis, and numerical simulations.
These results are shown in Fig.~\ref{fig:singular_perturbation}.
In the stationary regime, all three methods yield consistent results.
Furthermore, as the system approaches the bifurcation point, the results of the singular perturbation and the numerical simulations begin to deviate from those of the linear analysis.
The singular perturbation results successfully capture this deviation.
Moreover, even beyond the bifurcation point, the results of the singular perturbation analysis remain consistent with those of the numerical simulations.
This agreement is expected only in the vicinity of the bifurcation point, where $\epsilon$ is sufficiently small.
As the system moves farther into the oscillatory regime, deviations from the numerical results begin to appear.
\subsection{Deterministic limit of the learning rate}
\begin{figure}[htbp]
    \centering
    \includegraphics[width=80mm]{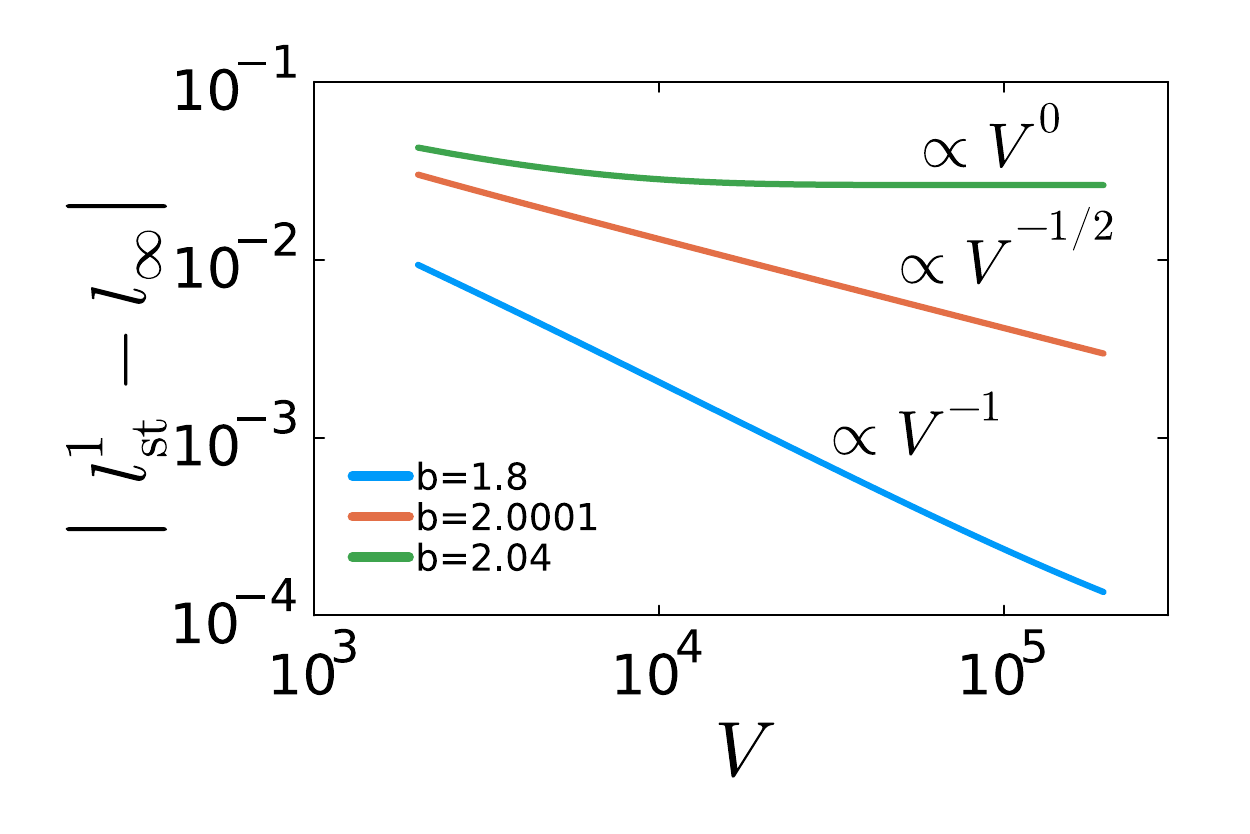}
    \caption{System-size dependence of the difference between the learning rate $l^1_{\mathrm{st}}$ and its asymptotic value in the stationary regime $l_\infty$.}
    \label{fig:convergence_systemsize}
\end{figure}

To analyze the singular behavior of the learning rate at the Hopf bifurcation, we investigate the deterministic limit $V\to\infty$ of the learning rate in the Brusselator.
In the oscillatory regime, as the noise vanishes in the limit $V\to\infty$, the stationary distribution becomes sharply localized around the deterministic limit cycle.
The limiting learning rate can then be interpreted as quantifying the contribution of the cyclic motion along this sharply attractor to changes in the correlation between the two variables.
\begin{figure}[htbp]
    \centering
    \includegraphics[width=80mm]{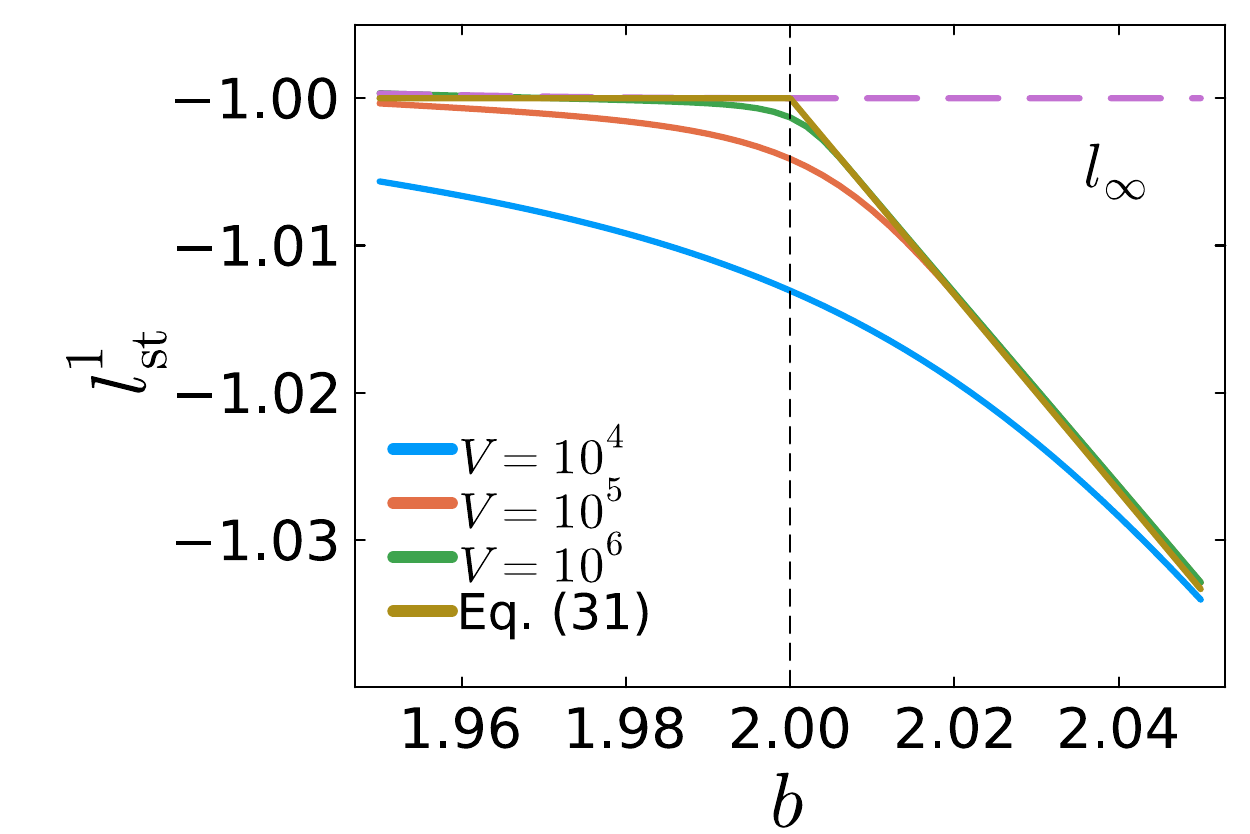}
    \caption{Deterministic limit of the learning rate $l^1_{\mathrm{st}}$ near the bifurcation point in the Brusselator with $b_c = 2$. The purple line $l_{\infty}$ represents the asymptotic value in the stationary regime, namely the result of the linear analysis extended into the oscillatory regime. The yellow line represents the learning rate in the deterministic limit gevin by \eqref{lst_piecewise}.}
    \label{fig:Brusselator_learning_limit}
\end{figure}
Using the analytical expression \eqref{transformation_learning}, we compute the learning rate near the bifurcation point while changing the system size.
The results are shown in Fig.~\ref{fig:Brusselator_learning_limit}.
As the system size increases, the learning rate in the stationary regime converges to a constant value.
In contrast, in the oscillatory regime, it asymptotically approaches a linearly decreasing trend with respect to the bifurcation parameter.
We analyze the convergence behavior of the learning rate.
Let $l_{\infty}$ denote  the asymptotic value in the stationary regime, namely the result of the linear analysis extended into the oscillatory regime.
Fig. \ref{fig:convergence_systemsize}  shows how the deviation from $l_{\infty}$ behaves as a function of the system size.
When the system resides in the stationary regime, the deviation from $l_\infty$ decays as $V^{-1}$.
At the bifurcation point, the decay becomes slower and follows a $V^{-1/2}$ scaling. 
In the oscillatory regime, the deviation converges to a finite value as $V \to \infty$.
These results demonstrate that the learning rate converges to distinct values in the stationary and oscillatory regimes, respectively.

We derive the learning rate in the deterministic limit.
Fixing $\epsilon$ as a small parameter and taking $V\to\infty$, the stationary distribution \eqref{stationary_distribution} becomes sharply concentrated.
This enables us to evaluate the learning rate in a systematic expansion in $\epsilon$.
For the Brusselator with $a=1$, we obtain the learning rate in the deterministic limit as
\begin{equation}
\label{lst_piecewise}
\begin{split}
\lim_{V\rightarrow \infty}
l^1_{\mathrm{st}}(\epsilon,V)
=
\begin{cases}
-1
+\mathcal{O}(\epsilon^3)
& \mathrm{for} \quad b \leq b_c,\\[1mm]
-1-\frac{4}{3}\epsilon^2
+\mathcal{O}(\epsilon^3)
& \mathrm{for} \quad b > b_c.
\end{cases}
\end{split}
\end{equation}
For the general expression, see Appendix~\ref{appendix_asymptotic_learning}.
We compare the theoretical deterministic limit \eqref{lst_piecewise} with numerical simulations in Fig.~\ref{fig:Brusselator_learning_limit}.
In the stationary regime ($b\leq b_c$), \eqref{lst_piecewise} coincides with the leading-order result of the linear analysis, and the numerical results converge to the same value as $V$ increases.
In the oscillatory regime ($b>b_c$), additional $\epsilon^2$ contributions appear in \eqref{lst_piecewise}.
This term yields a linear dependence on the bifurcation parameter $b$, so that the learning rate in the deterministic limit becomes a straight line with a finite slope.
These observations confirm that \eqref{lst_piecewise} is consistent with both the numerical simulations and the leading-order of the linear analysis.
As a result, the learning rate changes non-smoothly at the bifurcation point in the deterministic limit $V\to\infty$, and the limiting values on both sides of the bifurcation are evaluated analytically.

\section{Singular perturbation method}
\subsection{Stochastic normal form}
We outline the calculation based on a singular perturbation method.
The singular perturbation method is a theoretical method for expanding the equations and variables in terms of a small bifurcation parameter, where a coordinate transformation that simplifies the resulting equations is also expanded.
The simplified equation is referred to as a normal form.
In this paper, we extend the singular perturbation method to the Langevin equation \eqref{langevin} and derive a stochastic normal form.
For the singular perturbation analysis, we use notation based on \cite{Kuramoto1984}.

The time-evolution equation for $\vec{u}$ introduced in \eqref{intro_u} can be written as
\begin{equation}
\label{time_evolution_u}
    \frac{d}{dt}\vec{u}=
    L\vec{u}+M\vec{u}\vec{u}+N\vec{u}\vec{u}\vec{u}+\cdots
    +B(\vec{u}+\vec{x}_0)\cdot \vec{\xi}(t),
\end{equation}
where the term $M\vec{u}\vec{u}$ is explicitly written as
\begin{equation}
\begin{split}
(M\vec{u}\vec{u})_i
    =
    \sum_{j,k}\frac{1}{2!} \frac{\partial^2 F_i}{\partial x_j \partial x_k}\Big\vert_{\vec{x}=\vec{x}_0}u_j u_k,
\end{split}
\end{equation}
and the term $N\vec{u}\vec{u}\vec{u}$ is also written as
\begin{equation}
    (N\vec{u}\vec{u}\vec{u})_i
    =
    \sum_{j,k,l}\frac{1}{3!} \frac{\partial^3 F_i}{\partial x_j \partial x_k \partial x_l}\Big\vert_{\vec{x}=\vec{x}_0}u_j u_k u_l.
\end{equation}
Furthermore, we expand the linear operator introduced in \eqref{linear_operator} with respect to $\epsilon$ as
\begin{equation}
    L_{ij}
    =
    L_{0,ij}
    +
    \epsilon^2 \chi L_{1,ij}
    +
    \cdots.
\end{equation}
Because the system undergoes a Hopf bifurcation, $L_0$ has a pair of purely imaginary eigenvalues $\pm \mathrm{i}\omega_0$.
Let $\vec{U}$ and $\vec{U}^{\ast}$ denote the right and left eigenvectors of $L_0$ corresponding to the eigenvalue $\mathrm{i}\omega_0$. That is, we have
\begin{equation}
L_0 \vec{U}
=
\mathrm{i}\omega_0 \vec{U} \quad \mathrm{and} \quad
\vec{U}^{\ast} L_0
=
\mathrm{i}\omega_0 \vec{U}^{\ast}.
\end{equation}
The left and right eigenvectors are chosen to satisfy the normalization conditions:
\begin{equation}
\vec{U}^{\ast} \vec{U} =
\Bar{\vec{U}}^{\ast} \Bar{\vec{U}} = 1
\quad \mathrm{and} \quad
\vec{U}^{\ast} \Bar{\vec{U}} =
\Bar{\vec{U}}^{\ast} \vec{U} = 0,
\end{equation}
where the symbol $\bar{\cdot}$ represents the complex conjugate. 
Let $\vec{u}^{(0)}$ denote the solution to the linearized system at the bifurcation point.
We write
\begin{equation}
    \vec{u}^{(0)}
    =
    W\mathrm{e}^{\mathrm{i}\omega_0 t}\vec{U}+\mathrm{c.c.},
\end{equation}
where $W$ is an arbitrary complex number and $\mathrm{c.c.}$ represents the complex conjugate. Near the bifurcation point, the real part of the eigenvalues of the operator $L$ is of order $\epsilon^2$.
Because the amplitude of oscillation changes slowly on this timescale, we introduce a slow time variable $\tau \equiv \epsilon^2 t$ and assume that $W$ depends on $\tau$.
For later convenience, we set $\widetilde{W}(t,\tau)\equiv W(\tau) \mathrm{e}^{\mathrm{i}\omega_0 t}$.
Then, because the solution $\vec{u}$ is influenced by perturbations arising from the nonlinear terms and higher-order contributions in $\epsilon$, we assume the variable transformation between $\vec{u}$ and $\widetilde{W}$ as 
\begin{equation}
\begin{split}
\label{transformation_perturbation}
    \vec{u}&=\epsilon\vec{u}^{(1)}(\widetilde{W},\bar{\widetilde{W}})+\epsilon^2\vec{u}^{(2)}(\widetilde{W},\bar{\widetilde{W}})\\
    &+\epsilon^3\vec{u}^{(3)}(\widetilde{W},\bar{\widetilde{W}})+\cdots
\end{split}
\end{equation}
and the stochastic normal form as
\begin{equation}
\label{stochastic_normal_form}
    \frac{d}{dt} \widetilde{W}(t,\tau) = \mathrm{i}\omega_0 \widetilde{W}+\epsilon^2 A(\widetilde{W},\bar{\widetilde{W}}) + \eta'(t), 
\end{equation}
where 
\begin{equation}
\label{transformation_first}
\begin{split}
    \vec{u}^{(1)}
    &=
    \widetilde{W} (t,\tau)\vec{U}+\mathrm{c.c},
\end{split}
\end{equation}
and the functions $\vec{u}^{(2)}(\widetilde{W},\bar{\widetilde{W}})$, $\vec{u}^{(3)}(\widetilde{W},\bar{\widetilde{W}})$ and $A(\widetilde{W},\bar{\widetilde{W}})$ can be determined perturbatively.
We also assume that the variable transformation is determined solely by the deterministic part of \eqref{time_evolution_u}.
The noise term $\eta’(t)$ inherits the statistical properties of the noise in \eqref{time_evolution_u}.
As a result of the singular perturbation in Appendix \ref{appendix_perturbation}, $A(\widetilde{W},\bar{\widetilde{W}})$ can be obtained explicitly at the lowest order:
\begin{equation}
\label{normal_form}
A(\widetilde{W},\bar{\widetilde{W}})
=  \chi (\sigma_1+\mathrm{i}\omega_1) \widetilde{W} - (g_1 + \mathrm{i} g_2) |\widetilde{W}|^2 \widetilde{W} ,
\end{equation}
where $\sigma_1$, $\omega_1$, $g_1$ and $g_2$ are real constants independent of the bifurcation parameter $\mu$, or the small parameter $\epsilon$.
The noise term $\eta’$ satisfies the following statistical property:
\begin{equation}
\label{normal_noise_property}
    \langle
    \eta'(t_1) \bar{\eta}'(t_2)\rangle
    =
    \sum_{i,k} U_i^{\ast} \bar{U}_j^{\ast} \sum_{j} B^{(0)}_{ij} B^{(0)}_{kj}
    \delta(t_1-t_2),
\end{equation}
where we expand the original noise coefficient as
\begin{equation}
    B_{ij}(\vec{u}+\vec{x}_0)
    =
    \frac{1}{\sqrt{V}}(
    B^{(0)}_{ij}
    +
    \epsilon
    B^{(1)}_{ij}
    +
    \cdots),
\end{equation}
and define $B^{(0)}$ as the leading term with respect to the system size $V$ and the small parameter $\epsilon$ in the noise term.

The transformation is obtained as follows.
From the second-order equation in $\epsilon$, we obtain the second-order term of the coordinate transformation:
\begin{equation}
\label{transformation_second}
\begin{split}
    \vec{u}^{(2)}(\widetilde{W},\bar{\widetilde{W}})
    &=
    |\widetilde{W}|^2\vec{V}_0+(\widetilde{W})^2\vec{V}_+ +(\bar{\widetilde{W}})^2\vec{V}_-\\
    &+v_2\widetilde{W} \vec{U}+\bar{v_2}\bar{\widetilde{W}}\bar{\vec{U}},
\end{split}
\end{equation}
where the vectors $\vec{V}_0$, $\vec{V}_+$ and $\vec{V}_-$ are constant vectors independent of the system variables and the bifurcation parameter, and $v_2$ represents a coefficient that reflects the indeterminacy associated with the zero mode.
The second-order term in the variable transformation \eqref{transformation_second} constitutes the leading nonlinear correction to the neutrally stable solution \eqref{transformation_first}.
Finally, from the third-order equation in $\epsilon$, we obtain the third-order term of the coordinate transformation:
\begin{equation}
\label{transformation_third}
\begin{split}
    \vec{u}^{(3)}(\widetilde{W},\bar{\widetilde{W}})
    &=\chi \widetilde{W}\vec{h}_1
    +|\widetilde{W}|\widetilde{W}\vec{h}_{3,1}
    +(\widetilde{W})^3\vec{h}_{3,3}\\
    &+
    \chi \bar{\widetilde{W}}\bar{\vec{h}}_1
    +|\bar{\widetilde{W}}|\bar{\widetilde{W}}\bar{\vec{h}}_{3,1}
    +(\bar{\widetilde{W}})^3\bar{\vec{h}}_{3,3}\\
    &+
    v_2 (\widetilde{W})^2\vec{h}_{2}
    +\bar{v}_2 (\bar{\widetilde{W}})^2\bar{\vec{h}}_{2}
    +2v_2 |\widetilde{W}|^2 \vec{h}_{0}\\&+v_3\widetilde{W} \vec{U}+\bar{v_3}\bar{\widetilde{W}}\bar{\vec{U}},
\end{split}
\end{equation}
where $\vec{h}_1$, $\vec{h}_{3,1}$, $\vec{h}_{2}$ and $\vec{h}_{0}$ are constant vectors independent of the system variables and the bifurcation parameter.
$\vec{u}^{(3)}$ inherits the undetermined term in $\vec{u}^{(2)}$, and $v_3$ represents a coefficient that reflects the indeterminacy associated with the zero mode.

In the case of the Brusselator, the constant terms and vectors are determined as shown in Appendix \ref{Brusselator}.
Also, $v_2$ and $v_3$ represent the coefficients of the undetermined terms in the second and third order variable transformations, respectively.
In the present singular perturbation analysis, the expansion is carried out up to third order terms.
However, within this order of perturbation, the values of $v_2$ and $v_3$ could not be determined.

\subsection{Stationary distribution}
In this section, we derive the stationary distribution of the normal form coordinate, which was presented in \eqref{stationary_distribution} for large $V$.
For later convenience, we express the stochastic normal form in Cartesian coordinates as follows:
\begin{equation}
\label{def_y}
\widetilde{W} = y_1 + \mathrm{i} y_2,
\end{equation}
where $\boldsymbol{y} = (y_1, y_2) \in \mathbb{R}^2$ is the normal form coordinate.
The coordinate transformation $\vec{u}=\vec{f}(\boldsymbol{y})$ between $\boldsymbol{y}$ and the original variables $\vec{u}$ is obtained through \eqref{def_y} using \eqref{transformation_first} with Eqs. (\ref{transformation_perturbation}), (\ref{transformation_second}) and (\ref{transformation_third}) as
\begin{equation}
\label{transformation}
\begin{split}
    \vec{f}(\boldsymbol{y})&=\epsilon\vec{u}^{(1)}(y_1 + \mathrm{i} y_2,y_1 - \mathrm{i} y_2)+\epsilon^2\vec{u}^{(2)}(y_1 + \mathrm{i} y_2,y_1 - \mathrm{i} y_2)\\&+\epsilon^3\vec{u}^{(3)}(y_1 + \mathrm{i} y_2,y_1 - \mathrm{i} y_2)
\end{split}
\end{equation}

The Langevin equation for $\boldsymbol{y}$ is obtained from the stochastic normal form, and the drift term and diffusion matrix of the corresponding Fokker–Planck equation \eqref{normal_Fokker} are given as follows.
The drift term $\tilde{\boldsymbol{F}}(\boldsymbol{y})$ is obtained from Eqs. (\ref{stochastic_normal_form}) and (\ref{normal_form}) as
\begin{widetext}
\begin{equation}
\begin{split}
    \tilde{\boldsymbol{F}}(\boldsymbol{y})
    &=
    \left(
    \begin{matrix}
        \epsilon^2 \chi \sigma_1 y_1 -(\epsilon^2\chi \omega_1 +\omega_0)y_2\\
        (\epsilon^2\chi \omega_1 +\omega_0)y_1+ \epsilon^2 \chi \sigma_1 y_2 
    \end{matrix}
    \right)
    -
    \left(
    \begin{matrix}
        \epsilon^2(y_1^2+y_2^2)(g_1 y_1-g_2 y_2)\\
        \epsilon^2(y_1^2+y_2^2)(g_2 y_1+g_1 y_2)
    \end{matrix}
    \right)
\end{split}
\end{equation}
\end{widetext}
and the diffusion matrix $\tilde{D}$ is obtained from \eqref{normal_noise_property} as
\begin{widetext}
\begin{equation}
\label{normal_form_diffusion}
    \begin{split}
        \tilde{D}
        &=
        \frac{1}{4}
        \left(
        \begin{matrix}
            \sum_{i,k} (2U_i^{\ast} \bar{U}_j^{\ast}+U_i^{\ast}U_k^{\ast}+\bar{U}_i^{\ast}\bar{U}_k^{\ast}) \sum_{j} B^{(0)}_{ij} B^{(0)}_{kj}&\mathrm{i}\sum_{i,k}(\bar{U}_i^{\ast}\bar{U}_k^{\ast}-U_i^{\ast}U_k^{\ast})\sum_{j} B^{(0)}_{ij} B^{(0)}_{kj}\\
            \mathrm{i}\sum_{i,k}(\bar{U}_i^{\ast}\bar{U}_k^{\ast}-U_i^{\ast}U_k^{\ast})\sum_{j} B^{(0)}_{ij} B^{(0)}_{kj}&\sum_{i,k} (2U_i^{\ast} \bar{U}_j^{\ast}-U_i^{\ast}U_k^{\ast}-\bar{U}_i^{\ast}\bar{U}_k^{\ast}) \sum_{j} B^{(0)}_{ij} B^{(0)}_{kj}
        \end{matrix}
        \right).
    \end{split}
\end{equation}
\end{widetext}
Then, using the steady-state condition 
\begin{equation}
    \tilde{\boldsymbol{\nu}}_{\mathrm{st}} \cdot \partial_{\boldsymbol{y}} \ln \tilde{p}_\mathrm{st} + \partial_{\boldsymbol{y}} \cdot \tilde{\boldsymbol{\nu}}_{\mathrm{st}}=0,
\end{equation}
the stationary distribution is obtained as \eqref{stationary_distribution} in the regime where the bifurcation parameter and the system size satisfy $\epsilon \ll 1$ and $V \gg 1$. For details, see Appendix \ref{stationary}.

The statistical quantities in the original coordinate are obtained via the variable transformation \eqref{transformation} and  the stationary distribution \eqref{stationary_distribution}. The transformations Eqs. (\ref{transformation_second}) and (\ref{transformation_third}) involve the undetermined coefficients $v_2$ and $v_3$.
These coefficients cannot be fixed within a general framework of the singular perturbation analysis and depend on the observable of interest.
Therefore, because the goal of this study is to calculate the learning rate, we fix them such that the learning rate obtained via the singular perturbation method agrees with that from the linear analysis in the stationary regime:
\begin{equation}
    v_2=0,\quad v_3 = \frac{(3a^2+2)\chi}{2(2a^2+1)}.
\end{equation}
Using the Fokker-Planck equation for the normal form coordinate $\boldsymbol{y}$ given by \eqref{normal_Fokker} and the transformation from the normal form coordinate to the original variable $\vec{u}$ given by \eqref{transformation}, we calculate the learning rate of the original variables.
\subsection{Statistical quantity of the original coordinate}
To further demonstrate the validity of our approach, we compare the results of the singular perturbation analysis with those from numerical simulations for the steady-state variance of variable $\eta_1$, which is defined as
\begin{equation}
    \hat{\Xi}_{\mathrm{st},11} = V \int d \vec{x} p_{\mathrm{st}}(\vec{x}) (x_1 - \langle x_1\rangle)^2.
\end{equation}
In the stationary regime, $\hat{\Xi}_{\mathrm{st},11}$ can be calculated via the linear analysis as
\begin{equation}
\label{linear_variance}
    \hat{\Xi}_{\mathrm{st},11}
    =
    \frac{a(b_c+b)}{b_c-b}
\end{equation}
(see Appendix \ref{appendix_linear_learning}).
We compare three results for the variance: the result from the linear analysis, the theoretical prediction obtained using the variable transformation \eqref{transformation} and the stationary distribution in the normal form coordinate \eqref{stationary_distribution}, and the numerical simulation results. The theoretical prediction is given by
\begin{equation}
\label{perturbative_variance}
    \begin{split}
        \hat{\Xi}_{\mathrm{st},ij}
        &=
        V
        \int d \vec{x} (x_i-\langle x_i\rangle)(x_j-\langle x_j\rangle) p_{\mathrm{st}}(\vec{x})\\
        &=
        V
        \int d\boldsymbol{y} u_i(\boldsymbol{y})u_j(\boldsymbol{y}) \tilde{p}_{\mathrm{st}}(\boldsymbol{y}).
    \end{split}
\end{equation}
\begin{figure}[tbph]
    \centering
    \includegraphics[width=80mm]{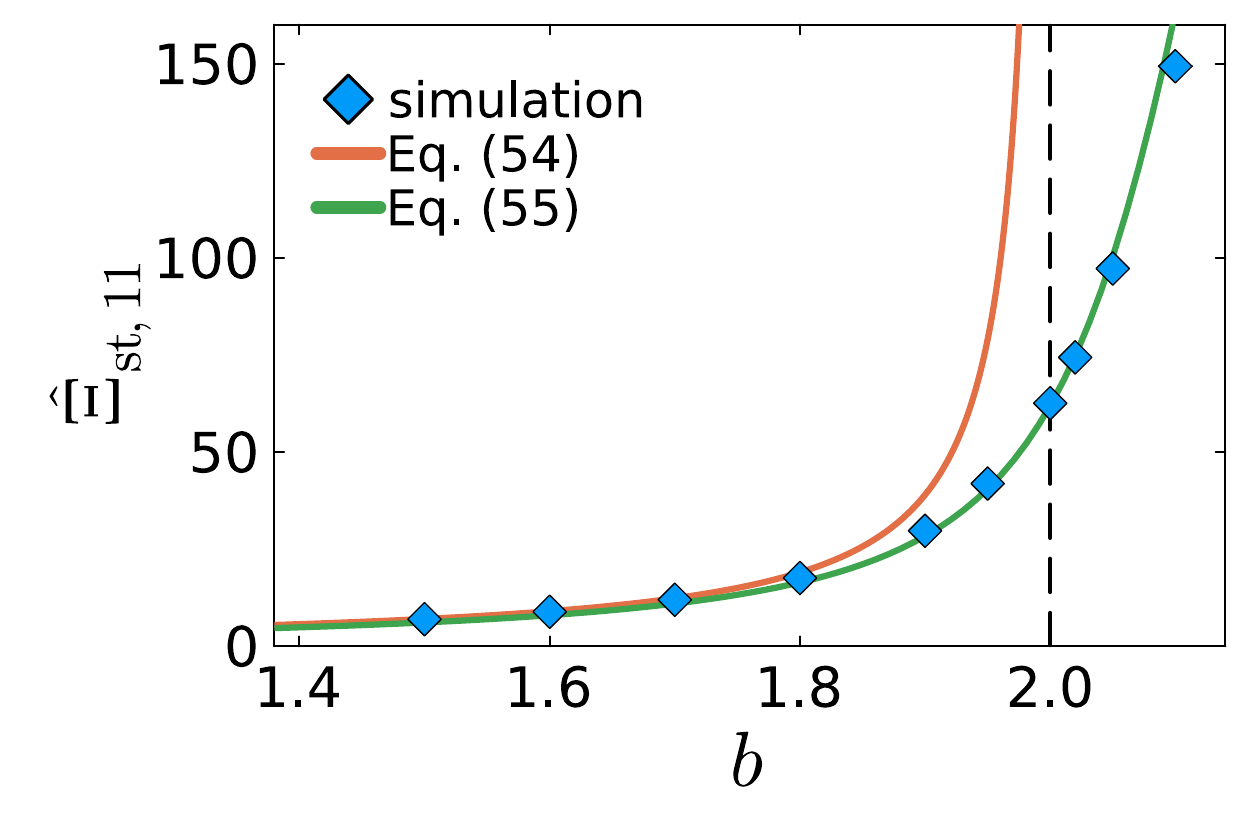}
    \caption{Variance of the variable $\eta_1$. $b_c = 2$ and $V = 2 \times 10^3$. In the stationary regime, the blue dots represent the results of numerical simulations, the orange line corresponds to the result from the linear analysis \eqref{linear_variance}, and the green line shows the theoretical prediction obtained via singular perturbation analysis \eqref{perturbative_variance}.}
    \label{fig:variance}
\end{figure}
The comparison is shown in Fig.~\ref{fig:variance}.
First, in the stationary regime, we observe that the three results are in agreement.
As the system approaches the bifurcation point, a discrepancy arises between the linear analysis and the numerical simulation.
However, the theoretical prediction based on singular perturbation analysis successfully reproduces the numerical simulation results.
Moreover, unlike the linear analysis, the singular perturbation method remains applicable beyond the bifurcation point, and its results continue to reproduce those of the numerical simulations.
This demonstrates that the singular perturbation method and the stationary distribution obtained in the normal form coordinate enable the reconstruction of the statistical properties in the original coordinates.
\section{Concluding remarks}
In this study, we have analyzed the behavior of the learning rate near a Hopf bifurcation, where the attractor changes from a stable fixed point to a limit cycle.
To this end, we extended the singular perturbation method—well-established in the study of deterministic dynamical systems—to the Langevin equation.
The novelty of the present analysis lies not in the general perturbative framework itself, but in carrying out the stochastic construction explicitly to the order required to evaluate the learning rate in the original variables.
This analysis revealed that the learning rate exhibits a non-smooth change at the bifurcation point.
Before ending this paper, we make a few remarks.

The singular perturbation method developed here is not limited to the Brusselator model but is applicable to a broad class of systems described by Langevin equations exhibiting a Hopf bifurcation.
Furthermore, this framework allows for the computation of not only the learning rate but also a variety of other statistical and thermodynamic quantities relevant to information thermodynamics.
Therefore, the analytical framework developed in this paper enables the computation of a wide range of statistical quantities that have been examined in previous studies of biochemical oscillators \cite{Stochastic_Thermodynamics_in_Mesoscopic,The_free-energy_cost, The_energy_cost_and_optimal_design,Improved_estimation_for_energy}.
In the present Brusselator model, the evolution equations are written in dimensionless form by setting the reaction-rate constants to unity.
In principle, the learning rate can be converted to physical units such as nats/s by restoring the original dimensional timescale.
A detailed discussion of its physical magnitude and dimensional interpretation is left for future work, since we focus here on its singular behavior at the Hopf bifurcation point.

Our standpoint is also different from recent studies that discuss information thermodynamics directly at the macroscopic or deterministic level, such as \cite{Freitas2023}. 
In the present work, we first define the learning rate for the underlying stochastic process and then investigate its deterministic limit. 
These two procedures are not generally equivalent. We adopt this stochastic-process-first approach because thermodynamic quantities and relations are well established for stochastic processes. 
In this sense, the present work should be viewed as a step toward a unified understanding of nonlinear dynamics, information-theoretic quantities, and thermodynamic structure, while the full interpretation in the deterministic limit remains to be clarified.

A key observation is that the learning rate remains nonzero even in the deterministic limit. 
This is made possible by reinterpreting the system as a stochastic process and then taking the vanishing noise limit.
In terms of its definition, the finite limiting value means that the evolution of each variable continues to change the correlation between the variables at a finite rate.
While this approach enables us to define information-theoretic quantities in deterministic systems, the actual values of these quantities still depend on the way noise is introduced. 
In the present study, the noise is not introduced arbitrarily, but is chosen to be consistent with that derived from the Kramers–Moyal expansion of the master equation.
Even with this physically motivated choice, the interpretation of these values as quantities intrinsic to the deterministic dynamics remains an open question.
A promising future direction is to identify forms of information flow that robustly reflect the intrinsic structure of deterministic dynamics, particularly in systems such as biochemical oscillators.

In this work, we focused on the connection between nonlinear dynamical behavior and information flow.
The learning rate, in particular, serves as a bridge linking the system’s information-theoretic and thermodynamic structures.
Our findings thus provide a foundation for a unified understanding of nonlinear phenomena, information theory, and thermodynamic structure.

\begin{acknowledgments}
This work was supported by JSPS KAKENHI (Grant Numbers JP22K13975, JP23K22415, JP25K00923,  JP25H01975 and JP26H00383) and JST SPRING (Grant Number JPMJSP2110).
\end{acknowledgments}
\section*{DATA AVAILABILITY}
The processed datasets and Julia scripts required to generate the plots in this manuscript are available on GitHub \cite{GithubData}.
The corresponding raw simulation data were not stored because of their large size, but they can be regenerated using the Julia scripts provided in the repository.
\appendix
\section{Derivation of (\ref{Brusselator_linear_learning})}
\label{appendix_linear_learning}
We derive the learning rate using the linear analysis.
The Fokker–Planck equation corresponding to the linear Langevin equation is given as
\begin{equation}
\partial_t \hat{p}_t(\vec{\eta}) = - \sum_{i=1}^2 \sum_{j=1}^2 \left( \hat{K}_{t,ij} \eta_j - \frac{1}{2}\hat{D}_{t,ij} \partial_{\eta_j}  \right) \hat{p}_t(\vec{\eta}), \tag{\ref{linear_Fokker}}
\end{equation}
where $\hat{\boldsymbol{K}}_t$ is the drift matrix and $\hat{D}_t$ is the diffusion matrix independent of the variables.
The local mean velocity is given by
\begin{equation}
    \nu_{t,i}(\vec{\eta})
    =
    \sum_{j=1}^2 \left( \hat{K}_{t,ij} \eta_j - \frac{1}{2}\hat{D}_{t,ij} \partial_{\eta_j} \ln \hat{p}_t(\vec{\eta})  \right) ,
\end{equation}
The linear Langevin system maintains a Gaussian distribution at all times if and only if the initial distribution is Gaussian:
\begin{equation}
    \hat{p}_t(\vec{\eta})
    =
    \frac{1}{2\pi\sqrt{ \det (\hat{\boldsymbol{\Xi}}_t)}}
    \exp\left[-\frac{1}{2}\vec{\eta}^{\mathrm{T}} (\hat{\boldsymbol{\Xi}}_t)^{-1} \vec{\eta}\right].
\end{equation}
This is because, when the initial distribution is Gaussian, all cumulants of order three or higher vanish at any time. By using the Langevin equation corresponding to the above probability distribution, the irreversible circulation \cite{irriversible_circulation}
\begin{equation}
    \hat{\alpha}_{t,ij}
    \equiv
    \frac{1}{2}
    \left\langle
    \dot{\eta}_i \eta_j- \dot{\eta}_j \eta_i
    \right\rangle
\end{equation}
is transformed into
\begin{equation}
    \hat{\alpha}_{t}
    =
    \frac{1}{2}
    \left(
    \hat{K}_t \hat{\Xi}_t - \hat{\Xi}_t \hat{K}_t^{\mathrm{T}}     
    \right).
\end{equation}
Considering the time evolution of the covariance matrix, we also obtain
\begin{equation}
    \begin{split}
        \frac{d}{dt}
        \hat{\Xi}_{t,ij}
        &=
        \int d\vec{\eta} \frac{\partial}{\partial t} \hat{p}_t(\vec{\eta}) \eta_i \eta_j\\
        &=
        \int d\vec{\eta}
        \sum_{k,l} \frac{\partial}{\partial \eta_k} 
        \left(
        \hat{K}_{t,kl} \eta_l - \frac{1}{2}\hat{D}_{t,kl} \frac{\partial}{\partial \eta_l}
        \right)\hat{p}_t(\vec{\eta})
        \eta_i \eta_j\\
        &=
        \sum_k\left( \hat{K}_{t,ik}\hat{\Xi}_{t,kj}+ \hat{\Xi}_{t,ik} \hat{K}_{t,kj}
        \right) + \hat{D}_{t,ij}.
    \end{split}
\end{equation}
The probability current $\boldsymbol{J}_t$ is expressed in terms of the covariance matrix and the irreversible circulation as
\begin{equation}
    \vec{\nu}_t
    =
    \left(
    -\frac{1}{2} \frac{d}{dt} \hat{\Xi}_t + \hat{\alpha}_t
    \right)
    \cdot
    \frac{\partial}{\partial \vec{\eta}} \hat{p}_t(\vec{\eta}).
\end{equation}
In the stationary state, because $d\hat{{\boldsymbol{\Xi}}}_{\mathrm{st}}/dt = 0$, the learning rate is given by
\begin{equation}
    \label{learning_circulation}
    \begin{split}
        \hat{l}^{i}_{\mathrm{st}}
        &=
        \int d\vec{\eta} \nu_{\mathrm{st},i} \hat{p}_t(\vec{\eta})\frac{\partial}{\partial \eta_i} \ln{\hat{p}_t(\vec{\eta})}\\
        &=
        \sum_j \hat{\alpha}_{\mathrm{st},ij} ((\hat{\boldsymbol{\Xi}}_{\mathrm{st}})^{-1})_{ij},
    \end{split}
\end{equation}
which shows that the learning rate is directly related to the irreversible circulation \(\hat{\alpha}_{\mathrm{st}}\) of the stationary probability current around the fixed point.
For the Brusselator, the irreversible circulation and the covariance matrix are given by
\begin{equation}
\label{linear_irreversible}
    \hat{\alpha}_{\mathrm{st}}
    =
    -\frac{2ab}{b_c-b}
    \left(
    \begin{matrix}
        0&1\\
        -1&0
    \end{matrix}
    \right),
\end{equation}
and
\begin{equation}
\label{linear_variance2}
    \hat{\boldsymbol{\Xi}}_{\mathrm{st}}
    =
    \left(
    \begin{matrix}
        \frac{a(b_c+b)}{b_c-b}&-\frac{2ab}{b_c-b}\\
        -\frac{2ab}{b_c-b}&\frac{b(b_c+b)}{a(b_c-b)}
    \end{matrix}
    \right).
\end{equation}
By using Eqs. (\ref{linear_irreversible}) and (\ref{linear_variance2}), we obtain \eqref{Brusselator_linear_learning}.

\section{Derivation of (\ref{transformation_perturbation}) and (\ref{stochastic_normal_form})}\noindent
\label{appendix_perturbation}
The systems exhibiting the Hopf bifurcation, such as the Brusselator, are described by nonlinear equations.
For deterministic dynamics, we analyze the dynamics using the singular perturbation method near the bifurcation point.
This approach yields both the normal form of the Hopf bifurcation, which describes the amplitude and phase of oscillations, and the transformation from the normal form coordinate to the original variables simultaneously \cite{Kuramoto1984}.
In this section, we extend the singular perturbation method to the Langevin equation.
Then, we directly apply the deterministic analysis to determine the transformation from the normal form coordinate to the original variables, while incorporating the effect of noise into the time evolution equation.
\subsection{Deterministic part}
We start by introducing bifurcation analysis for systems described by deterministic equations. In this study, the variable transformation is determined based on the analysis of the deterministic part.
Suppose that the time evolution is described as
\begin{equation}
\label{ODE}
    \frac{d}{dt}\vec{x}
    =
    \vec{F}(\vec{x}).
\end{equation}
For the unique fixed point $\vec{x}_0$ satisfying
\begin{equation}
\vec{F}(\vec{x}_0) = \vec{0},
\end{equation}
we introduce a new variable
\begin{equation}
\vec{u}
\equiv
\vec{x}
-
\vec{x}_0
\end{equation}
such that the fixed point is shifted to the origin.
By expanding the right-hand side with respect to $\vec{u}$, \eqref{ODE} is rewritten as
\begin{equation}
\frac{d}{dt}\vec{u}
=
L\vec{u}
+
M : \vec{u}\vec{u}
+
N : \vec{u}\vec{u}\vec{u}
+
\cdots,
\end{equation}
where $L$, $M$ and $N$ are defined as follows:
\begin{equation}
L_{ij}
=
\left.\frac{\partial F_i}{\partial x_j} \right|_{\vec{x}=\vec{x}_0},
\end{equation}
\begin{equation}
(M:\vec{u}\vec{u})_{i}
=
\sum_{j,k}\frac{1}{2!}
\left.\frac{\partial^2 F_i}{\partial x_j\partial x_k}\right|_{\vec{x}=\vec{x}_0}u_j u_k,
\end{equation}
\begin{equation}
(N : \vec{u}\vec{u}\vec{u})_{i}
=
\sum_{j,k,l}\frac{1}{3!}
\left.\frac{\partial^3 F_i}{\partial x_j\partial x_k \partial x_l}\right|_{\vec{x}=\vec{x}_0}u_j u_k u_l.
\end{equation}
By expanding them in terms of $\mu$ introduced in \eqref{bifurcation_parameter}, we write
\begin{equation}
\label{L_mu}
L
=
L_0
+
\mu L_1
+
\mu^2L_2
+
\cdots,
\end{equation}
\begin{equation}
\label{M_mu}
M
=
M_0
+
\mu M_1
+
\mu^2M_2
+
\cdots,
\end{equation}
\begin{equation}
\label{N_mu}
N
=
N_0
+
\mu N_1
+
\mu^2N_2
+
\cdots.
\end{equation}
Let $\vec{U}$ denote the right eigenvector satisfying
\begin{equation}
L_0 \vec{U}
=
\mathrm{i}\omega_0 \vec{U}.
\end{equation}
Similarly, we define the left eigenvector $\vec{U}^{\ast}$ by
\begin{equation}
\vec{U}^{\ast} L_0
=
\mathrm{i}\omega_0 \vec{U}^{\ast}.
\end{equation}
The left and right eigenvectors are chosen to satisfy the normalization conditions.
\begin{equation}
\vec{U}^{\ast} \vec{U} =
\Bar{\vec{U}}^{\ast} \Bar{\vec{U}} = 1,
\end{equation}
\begin{equation}
\vec{U}^{\ast} \Bar{\vec{U}} =
\Bar{\vec{U}}^{\ast} \vec{U} = 0.
\end{equation}
Because the present analysis focuses on the vicinity of the bifurcation point, we introduce a small parameter $\epsilon$ for the bifurcation parameter as 
\begin{equation}
\mu = \epsilon^2 \mathrm{sgn}(\mu) = \epsilon^2 \chi,
\end{equation}
where $\chi = \mathrm{sgn}(\mu)$.
We then rewrite Eqs. (\ref{L_mu}), (\ref{M_mu}) and (\ref{N_mu}) as
\begin{equation}
L
=
L_0
+
\epsilon^2 \chi L_1
+
\epsilon^4 L_2
+
\cdots,
\end{equation}
\begin{equation}
M
=
M_0
+
\epsilon^2 \chi M_1
+
\epsilon^4 M_2
+
\cdots,
\end{equation}
\begin{equation}
N
=
N_0
+
\epsilon^2 \chi N_1
+
\epsilon^4 N_2
+
\cdots.
\end{equation}
We also expand the variable $\vec{u}$ as
\begin{equation}
\vec{u}
=
\epsilon \vec{u}^{(1)}
+\epsilon^2 \vec{u}^{(2)}
+\epsilon^3 \vec{u}^{(3)}
+\cdots.
\end{equation}
Here, we introduce the following slow timescale $\tau$ 
\begin{equation}
\tau = \epsilon^2 t,
\end{equation}
because the real part of $(\vec{U}^{\ast},L\vec{U})$ is proportional to $\epsilon^2$.
Assuming that the variable $\vec{u}$ depends on both timescales $t$ and $\tau$, the time derivative is expressed as
\begin{equation}
\frac{d}{dt}
\rightarrow
\frac{\partial}{\partial t}
+
\epsilon^2\frac{\partial}{\partial \tau}.
\end{equation}
Substituting the above terms into the original equation gives
\begin{equation}
\begin{split}
&\left(
\frac{\partial}{\partial t}+\epsilon^2\frac{\partial}{\partial \tau}-L_0-\epsilon^2\chi L_1
\right)
(\epsilon \vec{u}^{(1)}+\epsilon^2 \vec{u}^{(2)}+\cdots)\\
&=\epsilon^2 M_0 : \vec{u}^{(1)}\vec{u}^{(1)} +\epsilon^3(2M_0 : \vec{u}^{(1)}\vec{u}^{(2)})+\mathcal{O}(\epsilon^4).
\end{split}
\end{equation}
The equations corresponding to each order of $\epsilon$ are collectively written as
\begin{equation}
\label{PDE}
\left(
\frac{\partial}{\partial t}-L_0
\right)\vec{u}^{(\nu)}
=
\vec{S}_{\nu},
\end{equation}
where $\vec{S}_{\nu}$ represents the terms of order $\epsilon^\nu$. Here are the expressions for the source terms $\vec{S}_\nu$ for $\nu = 1, 2, 3$:
\begin{equation}
\vec{S}_1
=
0,
\end{equation}
\begin{equation}
\vec{S}_2
=
M_0 : \vec{u}^{(1)} \vec{u}^{(1)},
\end{equation}
\begin{equation}
\vec{S}_3
=
-\left(
\frac{\partial}{\partial \tau}
-\chi L_1
\right)\vec{u}^{(1)}
+
2M_0 : \vec{u}^{(1)}\vec{u}^{(2)}
+
N_0 : \vec{u}^{(1)} \vec{u}^{(1)} \vec{u}^{(1)}.
\end{equation}
We now solve \eqref{PDE}.
We first notice 
\begin{equation}
    \left(
    \frac{\partial}{\partial t}-L_0
    \right)\vec{U}\mathrm{e}^{\mathrm{i}\omega_0 t}
    =
    0,
\end{equation}
which leads to the statement that the solution $\vec{u}^{(\nu)}$ to \eqref{PDE} exists only when the condition 
\begin{equation}
\label{solvability_condition}
    \int^T_0 \vec{U}^{\ast} \cdot \vec{S}_\nu \mathrm{e}^{-\mathrm{i}\omega_0 t} dt
    =
    0
\end{equation}
holds.
This condition is referred to as the solvability condition.
For later convenience, we introduce the notation
\begin{equation}
(\vec{\alpha},\vec{\beta})
=
\frac{1}{T}\int^{T}_0 \vec{\alpha} \cdot \vec{\beta} \mathrm{e}^{-\mathrm{i}\omega_0 t} dt
\end{equation}
for $\vec{\alpha}(t)$ and $\vec{\beta}(t)$, where $T = 2\pi/\omega_0$ is the period of the oscillation. The solvability condition \eqref{solvability_condition} is then expressed as
\begin{equation}
(\vec{U}^{\ast}, \vec{S}_{\nu}) = 0.
\end{equation}
In the argument below, we impose the solvability condition to have a consistent perturbative expansion.
First, we investigate the case $\nu = 1$. The unperturbed equation for this case
\begin{equation}
\left(
\frac{\partial}{\partial t} - L_0
\right)\vec{u}^{(1)} = 0
\end{equation}
has a neutrally stable solution given by
\begin{equation}
\label{first_order}
\vec{u}^{(1)} = W(\tau) \mathrm{e}^{\mathrm{i} \omega_0 t} \vec{U} + \Bar{W}(\tau) \mathrm{e}^{-\mathrm{i} \omega_0 t} \Bar{\vec{U}},
\end{equation}
where $W(\tau)$ is a slowly varying complex amplitude depending on the slow time scale $\tau$.
Next, we study the case $\nu = 2$. The corresponding equation is
\begin{equation}
\left(
\frac{\partial}{\partial t} - L_0
\right)\vec{u}^{(2)}
=
M_0 : \vec{u}^{(1)} \vec{u}^{(1)}.
\end{equation}
We substitute \eqref{first_order} into the right-hand side
\begin{equation}
\begin{split}
    &\left(
    \frac{\partial}{\partial t}-L_0
    \right)\vec{u}^{(2)}
    \\&=
    W(\tau)^2\mathrm{e}^{2\mathrm{i}\omega_0 t}M_0 : \vec{U} \vec{U}\\
    &+
    \Bar{W}(\tau)^2\mathrm{e}^{-2\mathrm{i}\omega_0 t}M_0 : \Bar{\vec{U}} \Bar{\vec{U}}
    +
    2
    |W(\tau)|^2M_0 : \vec{U} \Bar{\vec{U}}.
\end{split}
\end{equation}
Because the right-hand side contains components oscillating at different frequencies $2\omega_0$, $-2\omega_0$ and $0$, we expand $\vec{u}^{(2)}$ in terms of Fourier modes as
\begin{equation}
    \vec{u}^{(2)}
    =
    \sum_{l=-\infty}^{\infty}
    \vec{u}^{(2)(l)}(\tau) e^{il\omega_0 t}.
\end{equation}
The second-order correction $\vec{u}^{(2)}$ is determined by solving for each Fourier component individually. These are calculated as
\begin{equation}
    \vec{u}^{(2)(0)}
    =
    -2|W|^2L_0^{-1}M : \vec{U}\Bar{\vec{U}},
\end{equation}
\begin{equation}
    \vec{u}^{(2)(2)}
    =
    W^2(2\mathrm{i}\omega_0-L_0)^{-1}M : \vec{U}\vec{U},
\end{equation}
\begin{equation}
    \vec{u}^{(2)(-2)}
    =
    \Bar{W}^2(-2\mathrm{i}\omega_0-L_0)^{-1}M : \vec{U}\vec{U},
\end{equation}
\begin{equation}
    \vec{u}^{(2)(l)}
    =
    0 \quad (l \neq -2,0,2),
\end{equation}
where we define $\vec{V}_{+},\vec{V}_{-}$ and $\vec{V}_{0}$ as
\begin{equation}
\label{V_plus}
    \vec{V}_{+} \equiv
    (2i\omega_0-L_0)^{-1}M_0 : \vec{U}\vec{U}=\Bar{\vec{V}}_{-},
\end{equation}
\begin{equation}
\label{V_0}
    \vec{V}_0\equiv
    -2L_0^{-1}M_0 : \vec{U}\Bar{\vec{U}}.
\end{equation}
Using these notations, the solution $\vec{u}^{(2)}$ is expressed by
\begin{equation}
\label{second_order}
    \begin{split}
        \vec{u}^{(2)}
        &=
        |W|^2\vec{V}_0
        +
        W^2\mathrm{e}^{2\mathrm{i}\omega_0 t}\vec{V}_{+}\\
        &+
        \bar{W}^2\mathrm{e}^{-2\mathrm{i}\omega_0 t}\vec{V}_{-}
        +v_2W\mathrm{e}^{\mathrm{i}\omega_0 t}\vec{U}
        +\bar{v}_2\bar{W}\mathrm{e}^{-\mathrm{i}\omega_0 t}\bar{\vec{U}},
    \end{split}
\end{equation}
where $v_2$ and $\bar{v}_2$ are coefficients corresponding to homogeneous solutions that lie in the null space of the operator $(\partial/\partial t - L_0)$.\\
Finally, we move to the case $\nu = 3$:
\begin{equation}
\begin{split}
    \left(
    \frac{\partial}{\partial t}-L_0
    \right)\vec{u}^{(3)}
    &=
    -\left(
    \frac{\partial}{\partial \tau} 
    -\chi L_1
    \right)\vec{u}^{(1)}\\
    &+
    2M_0 : \vec{u}^{(1)}\vec{u}^{(2)}
    +
    N_0 : \vec{u}^{(1)} \vec{u}^{(1)} \vec{u}^{(1)}.
\end{split}
\end{equation}
By substituting the previously obtained Eqs. (\ref{first_order}) and (\ref{second_order}) into the equation and setting 
\begin{equation}
    \dot{W}(\tau)\equiv \frac{\partial}{\partial \tau}W(\tau) ,
\end{equation}
we obtain
\begin{widetext}
\begin{equation}
\label{third_order_equation}
    \begin{split}
        \left(
        \frac{\partial}{\partial t}-L_0
        \right)\vec{u}^{(3)}
        &=
        -\dot{W} \mathrm{e}^{\mathrm{i}\omega_0 t}\vec{U}
        -\dot{\Bar{W}} \mathrm{e}^{-\mathrm{i}\omega_0 t}\Bar{\vec{U}}
        +
        W \mathrm{e}^{\mathrm{i}\omega_0 t}\chi L_0\vec{U}
        +\Bar{W} \mathrm{e}^{-\mathrm{i}\omega_0 t}\chi L_0\Bar{\vec{U}}\\
        &
        +2W^3M_0 : \vec{U}\vec{V}_{+}\mathrm{e}^{3i\omega_0 t}
        +2|W|^2\Bar{W}M_0 : \vec{U}\vec{V}_{-}\mathrm{e}^{-\mathrm{i}\omega_0 t}
        +2|W|^2\Bar{W}M_0 : \vec{U}\vec{V}_{0}\mathrm{e}^{\mathrm{i}\omega_0 t}\\
        &
        +2|W|^2WM_0 : \Bar{\vec{U}}\vec{V}_{+}\mathrm{e}^{\mathrm{i}\omega_0 t}
        +2\Bar{W}^3M_0 : \Bar{\vec{U}}\vec{V}_{-}\mathrm{e}^{-3\mathrm{i}\omega_0 t}
        +2|W|^2\Bar{W}M_0 : \Bar{\vec{U}}\vec{V}_{0}\mathrm{e}^{-\mathrm{i}\omega_0 t}\\
        &
        +W^3N_0 : \vec{U}\vec{U}\vec{U}\mathrm{e}^{3\mathrm{i}\omega_0 t}
        +3|W|^2WN_0 : \vec{U}\vec{U}\Bar{\vec{U}}\mathrm{e}^{\mathrm{i}\omega_0 t}
        +3|W|^2\Bar{W}N_0 : \vec{U}\Bar{\vec{U}}\Bar{\vec{U}}\mathrm{e}^{-\mathrm{i}\omega_0 t}\\
        &+\Bar{W}^3N_0 : \Bar{\vec{U}}\Bar{\vec{U}}\Bar{\vec{U}}\mathrm{e}^{-\mathrm{i}\omega_0 t}.
    \end{split}
\end{equation}
\end{widetext}
Performing the Fourier expansion of $\vec{u}^{(3)}$ and determining each coefficient, we obtain
\begin{equation}
    \begin{split}
        \vec{u}^{(3)}
        &=
        \chi W e^{\mathrm{i}\omega_0 t}\vec{h}_1
        +|W|^2W e^{\mathrm{i}\omega_0 t}\vec{h}_{3,1}
        +W^3 e^{3\mathrm{i}\omega_0 t}\vec{h}_{3,3}
        \\&+v_2 W^2 e^{2\mathrm{i}\omega_0 t}\tilde{\vec{h}}_{2}
        +v_2 |W|^2 \tilde{\vec{h}}_{0}
        +v_3 W e^{\mathrm{i}\omega_0 t}\vec{U} +\mathrm{c.c.}
    \end{split}
\end{equation}
with
\begin{equation}
    \begin{split}
        &\vec{h}_1
        =
        G(- \lambda_1 \vec{U}+L_1 \vec{U}),
        \\
        &\vec{h}_{3,1}
        =
        G\big(
        g\vec{U}
        +2M_0 : \vec{U}\vec{V}_{0}
        +2M_0 : \Bar{\vec{U}}\vec{V}_{+}
        +3N_0 : \vec{U}\vec{U}\Bar{\vec{U}}
        \big),
        \\
        &\vec{h}_{3,3}
        =
        (3\mathrm{i}\omega_0-L_0)^{(-1)}
        \left(
        2M_0 : \vec{U}\vec{V}_+ +N_0 : \vec{U}\vec{U}\vec{U}
        \right),
        \\
        &\tilde{\vec{h}}_{2}
        =
        2(2\mathrm{i}\omega_0-\vec{L}_0)^{-1}M_0 : \vec{U}\vec{U}
        =
        2\vec{V}_+,
        \\
        &\tilde{\vec{h}}_{0}
        =
        -2L_0^{-1}M_0 : \vec{U}\bar{\vec{U}}=\vec{V}_0,
    \end{split}
\end{equation}
where $G$ is the pseudo-inverse of $(\mathrm{i}\omega_0 - L_0)$, and $v_3$ is the coefficient corresponding to the undetermined term. 
From the solvability condition \eqref{solvability_condition}, we obtain 
\begin{equation}
    \begin{split}
        0&=
        -\dot{W}+\chi\lambda_1 W-g|W|^2W,
    \end{split}
\end{equation}
which leads to the evolution equation for the complex amplitude $W(\tau)$: 
\begin{equation}
    \dot{W}=\chi\lambda_1 W-g|W|^2W.
\end{equation}
Thus, we derive the evolution equation for the complex amplitude $W(\tau)$. Here,
\begin{equation}
\label{lambda_1}
    \lambda_1=(\vec{U}^{\ast}, L_1 \vec{U})
    \equiv \sigma_1+\mathrm{i}\omega_1
\end{equation}
and
\begin{equation}
\label{third_g}
\begin{split}
    g
    &= 
    (\vec{U}^{\ast},[-3N : \vec{U}_0\vec{U}_0\bar{\vec{U}}_0-2M : \vec{U}_0\vec{V}_0-2M : \bar{\vec{U}}_0\vec{V}_{+}])\\
    &\equiv g_1+\mathrm{i}g_2,
\end{split}
\end{equation}
where $\sigma_1$, $\omega_1$, $g_1$ and $g_2$ are real variables.
Finally, performing the transformation $W \mathrm{e}^{\mathrm{i} \omega_0 t} = y_1 + \mathrm{i} y_2$, we rewrite the time evolution equation as
\begin{equation}
\label{standard}
\begin{split}
    \frac{d}{dt} W\mathrm{e}^{\mathrm{i}\omega_0 t}
    &= (\epsilon^2 \chi\sigma + i\omega_0)We^{i\omega_0 t} 
    - \epsilon^2 g|W\mathrm{e}^{\mathrm{i}\omega_0 t}|^2W \mathrm{e}^{\mathrm{i}\omega_0 t},
\end{split}    
\end{equation}
which yields the time evolution equation in real orthogonal coordinates
\begin{widetext}
\begin{equation}
\label{normal form}
    \frac{d}{dt}
    \left(
    \begin{matrix}
    y_1\\
    y_2
    \end{matrix}
    \right)
    =
    \left(
    \begin{matrix}
        \epsilon^2 \chi\sigma_1 y_1 - (\epsilon^2\chi \sigma_2+ \omega_0) y_2 - \epsilon^2(y_1^2+y_2^2)(g_1 y_1 -  g_2 y_2)\\
        (\epsilon^2\chi \sigma_2+ \omega_0) y_1 + \epsilon \sigma_1\chi y_2 - \epsilon^2(y_1^2+y_2^2)(g_2 y_1 + g_1 y_2)
    \end{matrix}
    \right).
\end{equation}
\end{widetext}
The transformation to the original variables is determined by \eqref{transformation} with Eqs. (\ref{transformation_first}), (\ref{transformation_second}) and (\ref{transformation_third}).
As shown above, both the normal form and the corresponding variable transformation can be derived simultaneously using the singular perturbation method.
In this paper, when computing statistical quantities including the learning rate in the original coordinates, we use the variable transformation determined from the deterministic part.
\subsection{Noise part}
In this section, we perform a similar analysis for the Langevin equation. Because the drift term, which determines the deterministic behavior, can be treated in the same way, we focus on how to handle the noise.
Concretely, we expand the noise coefficient $B_t$ appearing in the Langevin equation with respect to the previously introduced small parameter $\epsilon$:
\begin{equation}
    B_{ij}(\vec{u}+\vec{x}_0)
    =
    \frac{1}{\sqrt{V}}(
    B^{(0)}_{ij}
    +
    \epsilon
    B^{(1)}_{ij}
    +
    \cdots),
\end{equation}
where the leading term $B^{(0)}_{ij}$ is independent of the variables $\vec{u}$. We assume that the leading term of the noise appears in the third-order equation.\\

We discuss the slow timescale $\tau$ in more detail. It should be noted that the following argument is not mathematically rigorous. We perform a trivial decomposition of time $t$ using an integer $n$ and the period $T$ as
\begin{equation}
t = nT + t’,\quad \mathrm{for} \quad t’ \in [0, T).
\end{equation}
Here, we define a time sequence $\tau_n$ as
\begin{equation}
\tau_n = \epsilon^2 n T.
\end{equation}
The interval of this time sequence is given by
\begin{equation}
\Delta \tau = \epsilon^2 T,
\end{equation}
which vanishes in the limit $\epsilon \rightarrow 0$. That is, the time sequence $\tau_n$ becomes continuous in the limit $\epsilon \rightarrow 0$. $\tau$ introduced in the previous section is the continuous time obtained in this way. Also, the integration over one period in the definition of the inner product can be understood as an integral with respect to $t’$. Keeping this treatment of the time scale in mind, let us examine the statistical properties of the term $\left(\vec{U}^{\ast},B^{(0)}\vec{\xi}\right)$, which is expected to appear in the stochastic normal form equation:
\begin{widetext}
\begin{equation}
    \begin{split}
        \left\langle
        \left(\vec{U}^{\ast},B^{(0)}\vec{\xi}\right)
        \times
        \left(\overline{\vec{U}^{\ast},B^{(0)}\vec{\xi}}\right)
        \right\rangle
        &=
        \Bigg\langle
        \left(
        \frac{1}{T}\int^T_0 dt'_1 \sum_i U_i^{\ast} \sum_{j} B^{(0)}_{ij} \xi_{j}(n_1,t'_1)\mathrm{e}^{-\mathrm{i}\omega_0 t'_1}
        \right)\\
        &\times
        \left(
        \frac{1}{T}\int^T_0 dt'_2 \sum_k \bar{U}_k^{\ast} \sum_{l} B^{(0)}_{kl} \xi_{l}(n_2,t'_2)\mathrm{e}^{\mathrm{i}\omega_0 t'_2}
        \right)
        \Bigg\rangle\\
        &=
        \frac{1}{T^2}
        \sum_{i,k} U_i^{\ast} \bar{U}_j^{\ast} \sum_{j,l} B^{(0)}_{ij} B^{(0)}_{kl}
        \int^T_0 dt'_1 \int^T_0 dt'_2
        \langle \xi_{j}(n_1,t'_1)\xi_{l}(n_2,t'_2) \rangle \mathrm{e}^{-\mathrm{i}\omega_0(t'_1-t'_2)}.
    \end{split}
\end{equation}
\end{widetext}
Here, since $\xi_{j}(n,t’)$ is white noise, its properties remain unchanged even on the macroscopic time scale. Therefore, the correlation transforms as
\begin{equation}
    \langle \xi_{j}(n_1,t'_1)\xi_{l}(n_2,t'_2) \rangle
    \rightarrow
    \delta_{j,l} \delta_{n_1,n_2} \delta(t'_1-t'_2).
\end{equation}
Using the white noise correlation, we obtain
\begin{equation}
\label{noise_times}
    \begin{split}
        &\left\langle
        \left(\vec{U}^{\ast},B^{(0)}\vec{\xi}\right)
        \times
        \left(\overline{\vec{U}^{\ast},B^{(0)}\vec{\xi}}\right)
        \right\rangle\\
        &=
        \frac{1}{T}
        \sum_{i,k} U_i^{\ast} \bar{U}_j^{\ast} \sum_{j,l} B^{(0)}_{ij} B^{(0)}_{kl}
        \delta_{j,l} \delta_{n_1,n_2}\\
        &=
        \frac{1}{T}
        \sum_{i,k} U_i^{\ast} \bar{U}_j^{\ast} \sum_{j} B^{(0)}_{ij} B^{(0)}_{kj}
        \delta_{n_1,n_2}.
    \end{split}
\end{equation}
From the argument on the time discretization, we replace the Kronecker delta over $n$ as
\begin{equation}
    \delta_{n_1,n_2}
    \rightarrow
    \epsilon^2 T \delta(\tau_1-\tau_2)
\end{equation}
in the limit $\epsilon \rightarrow 0$.
Substituting this into \eqref{noise_times} yields
\begin{equation}
\begin{split}
&\left\langle
\left(\vec{U}^{\ast},B^{(0)}\vec{\xi}\right)
        \times\left(\overline{\vec{U}^{\ast},B^{(0)}\vec{\xi}}\right)
        \right\rangle\\
        &=
    \epsilon^2
    \sum_{i,k} U_i^{\ast} \bar{U}_j^{\ast} \sum_{j} B^{(0)}_{ij} B^{(0)}_{kj}
    \delta(\tau_1-\tau_2).
\end{split}
\end{equation}
To simplify the notation, we introduce a new noise variable $\eta(\tau)$ defined by
\begin{equation}
\label{tau_noise_statistics}
    \langle
    \eta(\tau_1)\bar{\eta}(\tau_2)\rangle
    =
    \sum_{i,k} U_i^{\ast} \bar{U}_j^{\ast} \sum_{j} B^{(0)}_{ij} B^{(0)}_{kj}
    \delta(\tau_1-\tau_2).
\end{equation}
This noise $\eta(\tau)$ effectively captures the projection of the original noise onto the zero mode. With the above setup, it is now evident that the leading-order noise term appears at $\mathcal{O}(\epsilon^3)$ in the expansion, which is the same order at which the normal form is derived.\\

Combining the previous results from the deterministic analysis with the noise contribution, we obtain the normal form of the Hopf bifurcation with noise:
\begin{equation}
    \dot{W}(\tau)
    =
    \chi \sigma W(\tau) -g|W(\tau)|^2W(\tau) +\frac{1}{\epsilon^2\sqrt{V}} \eta(\tau).
\end{equation}
This equation describes the stochastic dynamics of the complex amplitude $W(\tau)$ near the Hopf bifurcation point, including both the nonlinear deterministic terms and the leading-order noise effect.
To use a variable transformation that does not explicitly depend on time, it is necessary to derive the time evolution equation for $W\mathrm{e}^{\mathrm{i}\omega_0 t}$. We can express the system in real-valued coordinate $(y_1, y_2)$ through the relation
$W\mathrm{e}^{\mathrm{i}\omega_0 t} = y_1 + \mathrm{i}y_2$,
which leads to the stochastic normal form enabling analysis of statistical quantities and learning rates in the original coordinates. Accordingly, the noise term $\eta’(t)$ that appears in the time evolution equation of $W\mathrm{e}^{\mathrm{i}\omega_0 t}$ is defined to satisfy the following statistical property:
\begin{equation}
\label{dash_noise_statistics}
    \langle
    \eta'(t_1) \bar{\eta}'(t_2)\rangle
    =
    \sum_{i,k} U_i^{\ast} \bar{U}_j^{\ast} \sum_{j} B^{(0)}_{ij} B^{(0)}_{kj}
    \delta(t_1-t_2).
\end{equation}
This ensures that the stochastic term in the transformed equation correctly reflects the noise correlations inherited from the original Langevin equation in real time $t$. The Langevin equation in real orthogonal coordinates under the variable transformation can be written as
\begin{widetext}
    \begin{equation}
    \frac{d}{dt}
    \left(
    \begin{matrix}
        y_1\\y_2
    \end{matrix}
    \right)
    =
    \left(
    \begin{matrix}
        \epsilon^2 \chi \sigma_1 y_1 -(\epsilon^2\chi \omega_1 +\omega_0)y_2
        -\epsilon^2(y_1^2+y_2^2)(g_1 y_1-g_2 y_2)\\
        (\epsilon^2\chi \omega_1 +\omega_0)y_1+ \epsilon^2 \chi \sigma_1 y_2 
        -\epsilon^2(y_1^2+y_2^2)(g_2 y_1+g_1 y_2)
    \end{matrix}
    \right)
    +
    \frac{1}{\epsilon\sqrt{V}}
    \left(
    \begin{matrix}
        \mathrm{Re}(\eta')\\
        \mathrm{Im}(\eta')
    \end{matrix}
    \right).
\end{equation}
\end{widetext}
This Langevin equation represents a normal form of the Hopf bifurcation with noise. The corresponding Fokker–Planck equation is given by \eqref{normal_Fokker}.
\section{Derivation of (\ref{stationary_distribution})}\noindent
\label{stationary}
In this section, we derive \eqref{stationary_distribution}, which represents the stationary distribution of the stochastic normal form near the Hopf bifurcation point.
Let $\tilde{p}_t(\boldsymbol{y})$ denote the probability distribution of $\boldsymbol{y}$. Then, the Fokker–Planck equation corresponding to the Langevin equation is given by
\begin{equation}
    \frac{\partial}{\partial t} \tilde{p}_t(\boldsymbol{y})
    =
    -\frac{\partial}{\partial \boldsymbol{y}}\cdot
    \left(
    \tilde{\boldsymbol{F}}(\boldsymbol{y})
    -\frac{1}{2\epsilon^2V}\tilde{D} \frac{\partial}{\partial \boldsymbol{y}}
    \right) \tilde{p}_t(\boldsymbol{y}),
\end{equation}
where $\tilde{\boldsymbol{F}}(\boldsymbol{y})$ is the drift vector and $\tilde{D}$ is the diffusion matrix in the stochastic normal form. Here, the drift $\tilde{\boldsymbol{F}}(\boldsymbol{y})$ is given by
\begin{widetext}
\begin{equation}
\begin{split}
    \tilde{\boldsymbol{F}}(\boldsymbol{y})
    &=
    \left(
    \begin{matrix}
        \epsilon^2 \chi \sigma_1 y_1 -(\epsilon^2\chi \omega_1 +\omega_0)y_2\\
        (\epsilon^2\chi \omega_1 +\omega_0)y_1+ \epsilon^2 \chi \sigma_1 y_2 
    \end{matrix}
    \right)
    -
    \left(
    \begin{matrix}
        \epsilon^2(y_1^2+y_2^2)(g_1 y_1-g_2 y_2)\\
        \epsilon^2(y_1^2+y_2^2)(g_2 y_1+g_1 y_2)
    \end{matrix}
    \right)
\end{split}
\end{equation}
\end{widetext}
and the diffusion matrix $\tilde{D}$ is given by
\begin{widetext}
\begin{equation}
\label{normal_form_diffusion}
    \begin{split}
        \tilde{D}
        &=
        \frac{1}{4}
        \left(
        \begin{matrix}
            \sum_{i,k} (2U_i^{\ast} \bar{U}_k^{\ast}+U_i^{\ast}U_k^{\ast}+\bar{U}_i^{\ast}\bar{U}_k^{\ast}) \sum_{j} B^{(0)}_{ij} B^{(0)}_{kj}&\mathrm{i}\sum_{i,k}(\bar{U}_i^{\ast}\bar{U}_k^{\ast}-U_i^{\ast}U_k^{\ast})\sum_{j} B^{(0)}_{ij} B^{(0)}_{kj}\\
            \mathrm{i}\sum_{i,k}(\bar{U}_i^{\ast}\bar{U}_k^{\ast}-U_i^{\ast}U_k^{\ast})\sum_{j} B^{(0)}_{ij} B^{(0)}_{kj}&\sum_{i,k} (2U_i^{\ast} \bar{U}_k^{\ast}-U_i^{\ast}U_k^{\ast}-\bar{U}_i^{\ast}\bar{U}_k^{\ast}) \sum_{j} B^{(0)}_{ij} B^{(0)}_{kj}
        \end{matrix}
        \right).
    \end{split}
\end{equation}
\end{widetext}
We now determine the stationary distribution of the Fokker–Planck equation \eqref{normal_Fokker}, which includes two small parameters: $\epsilon$, associated with the bifurcation parameter, and $1/V$, associated with the noise intensity.
We first assume the stationary distribution takes the asymptotic form
\begin{equation}
\label{distribution_form}
\tilde{p}(\boldsymbol{y}) = C_0 \exp\left[ V \phi(\boldsymbol{y})\right] \quad \mathrm{for} \quad V\gg 1,
\end{equation}
where $C_0$ is the normalization constant, and  $\phi(\boldsymbol{y})$ is a polynomial function of $\boldsymbol{y}$ that is independent of $V$.
By substituting \eqref{distribution_form} into the Fokker–Planck equation \eqref{normal_Fokker} and collecting terms order by order in $V$, we obtain
\begin{widetext}
\begin{equation}
\begin{split}
\label{fokker_order}
\tilde{\boldsymbol{F}}(\boldsymbol{y})\cdot\frac{\partial}{\partial \boldsymbol{y}}\phi(\boldsymbol{y})+\frac{1}{2\epsilon^2}\left(\frac{\partial}{\partial \boldsymbol{y}}\phi(\boldsymbol{y})\right)^{\mathsf T}\tilde{D}\left(\frac{\partial}{\partial \boldsymbol{y}}\phi(\boldsymbol{y})\right)=0.
\end{split}
\end{equation}
\end{widetext}
We then expand this function in powers of $\epsilon$ as
\begin{equation}
\phi(\boldsymbol{y})
=
\phi_0(\boldsymbol{y})
+
\epsilon^2\phi_1(\boldsymbol{y})
+
\epsilon^4\phi_2(\boldsymbol{y})
+
\mathcal{O}(\epsilon^6),
\end{equation}
By substituting this expansion into \eqref{fokker_order} and collecting terms of each order in $\epsilon$, we determine the function $\phi(\boldsymbol{y})$ by treating the result as an identity in $(y_1, y_2)$.
The function $\phi(\boldsymbol{y})$ is obtained as
\begin{equation}
    \begin{split}
    \phi(\boldsymbol{y})
    &= \epsilon^4  \left(
    \frac{2 \chi \sigma_1}{\tilde{D}_{11} + \tilde{D}_{22}} (y_1^2 + y_2^2)
    - \frac{g_1}{\tilde{D}_{11} + \tilde{D}_{22}} (y_1^2 + y_2^2)^2
    \right)\\
    &+\mathcal{O}(\epsilon^6),
    \end{split}
\end{equation}
where the constant terms are absorbed into the normalization constant $C_0$.
By extracting the leading-order terms in $\epsilon$ from  $\phi(\boldsymbol{y})$, we recover \eqref{stationary_distribution}.

\section{Derivation of (\ref{transformation_learning})}\noindent
\label{appendix_learning_transformation}
In this section, we derive \eqref{transformation_learning}. Let us start with the Fokker–Planck equation \eqref{Fokker_Planck}. In the steady state, applying integration by parts, the learning rate $l^i_{\mathrm{st}}$ for each variable $X_i$ is given by
\begin{equation}
\begin{split}
    l^i_{\mathrm{st}}
    &=
    \int d\vec{x} \nu_{t,i}(\vec{x})p(\vec{x})
    \frac{\partial}{\partial x_i}
    \ln{p(\vec{x})}.
\end{split}
\end{equation}
Next, we consider the case in which the variable $\vec{x}$ is reconstructed from another variable $\boldsymbol{y}$ via a transformation of the form $\vec{x} = \vec{f}(\boldsymbol{y})$. Then, we introduce the Jacobian matrix $R$ as
\begin{equation}
\label{jacobian}
    R_{ij}(\boldsymbol{y})
    =
    \frac{\partial}{\partial y_j} f_i(\boldsymbol{y}),
\end{equation}
where each element $R_{ij}$ represents the partial derivative of the $i$-th component of $\vec{x} = \vec{f}(\boldsymbol{y})$ with respect to $y_j$. In this case, the steady-state local probability current and probability density of the variable $X_i$ can be expressed in terms of the probability density $\tilde{p}(\boldsymbol{y})$ as
\begin{equation}
    p(\vec{f}(\boldsymbol{y}))=
    \frac{1}{\vert\det(R(\boldsymbol{y}))\vert}
    \tilde{p}(\boldsymbol{y})
\end{equation}
and the local mean velocity $\tilde{\boldsymbol{\nu}}(\boldsymbol{y})$ in the new variables $\boldsymbol{y}$ as
\begin{equation}
    \boldsymbol{\nu}(\vec{f}(\boldsymbol{y}))
    =
    R(\boldsymbol{y})
    \tilde{\boldsymbol{\nu}}(\boldsymbol{y}).
\end{equation}
Therefore, when expressed in terms of the original variables, the learning rate in the new variables becomes \eqref{transformation_learning}.

\section{Asymptotic behavior}
\label{appendix_asymptotic_learning}
We consider the asymptotic behavior of \eqref{transformation_learning} in the limit $V\rightarrow \infty$.
The learning rate \eqref{transformation_learning} is rewritten as
\begin{widetext}
\begin{equation}
\label{transformation_learning_asymptotic}
\begin{split}
    l^{i}_{\mathrm{st}}
    &=
    \int d\boldsymbol{y}\,
    \sum_{k,l}
    R_{ik}(\boldsymbol{y})\,
    \tilde{\nu}_{\mathrm{st},k}(\boldsymbol{y})\,
    \frac{\partial}{\partial y_l}
    \left[
    \ln \tilde{p}_{\mathrm{st}}(\boldsymbol{y})
    -
    \ln \big|\det R(\boldsymbol{y})\big|
    \right]\,
    R^{-1}_{li}(\boldsymbol{y})\,
    \tilde{p}_{\mathrm{st}}(\boldsymbol{y})\\
    &\equiv l^i_p+l^i_R.
\end{split}
\end{equation}
\end{widetext}

For convenience in the subsequent calculations, we write the stationary distribution \eqref{stationary_distribution} in the form
\begin{equation}
\label{asymptotic_stationary}
\begin{split}
    \tilde{p}_{\mathrm{st}}(y_1,y_2)
    &\equiv
    C_0
    \exp \Bigg\{
    \epsilon^4 V \left[
    \chi \alpha(y_1^2+y_2^2)+\beta (y_1^2+y_2^2)^2
    \right]
    \Bigg\},
\end{split}
\end{equation}
where $C_0$ is the normalization constant.
Furthermore, we perform an expansion with respect to $\epsilon$ for the quantities in \eqref{transformation_learning_asymptotic}.
The local mean velocity is expanded as
\begin{equation}
\label{asymptotic_local_mean}
\tilde{\boldsymbol{\nu}}_{\mathrm{st}}
=
\tilde{\boldsymbol{\nu}}^{(0)}_{(1)}(\boldsymbol{y})
+
\epsilon^2 \tilde{\boldsymbol{\nu}}^{(2)}_{(1)}(\boldsymbol{y})
+
\epsilon^2 \tilde{\boldsymbol{\nu}}^{(2)}_{(3)}(\boldsymbol{y}),
\end{equation}
where the superscript indicates the power of $\epsilon$, and the subscript denotes the order of $\boldsymbol{y}$.

We rewrite the Jacobian matrix \eqref{jacobian} as an expansion in $\epsilon$ as
\begin{widetext}
\begin{equation}
\label{asymptotic_jacobian}
R
=
\epsilon R^{(1)}_{(0)}
+\epsilon^2\Big(R^{(2)}_{(0)}(\boldsymbol{y})+R^{(2)}_{(1)}(\boldsymbol{y})\Big)
+\epsilon^3\Big(R^{(3)}_{(0)}(\boldsymbol{y})+R^{(3)}_{(1)}(\boldsymbol{y})+R^{(3)}_{(2)}(\boldsymbol{y})\Big)
+\mathcal{O}(\epsilon^4).
\end{equation}
\end{widetext}
We assume that the leading-order Jacobian $R^{(1)}_{(0)}$ is invertible.
By imposing $RR^{-1}=\boldsymbol{I}$ and matching powers of $\epsilon$,
the inverse expansion takes the form
\begin{widetext}
\begin{equation}
\begin{split}
\label{asymptotic_jacobian_inv}
R^{-1}
&=
\frac{1}{\epsilon}R^{(-1)}_{\mathrm{inv}(0)}
+
\Big(R^{(0)}_{\mathrm{inv}(0)}(\boldsymbol{y})+R^{(0)}_{\mathrm{inv}(1)}(\boldsymbol{y})\Big)
+\epsilon\Big(R^{(1)}_{\mathrm{inv}(0)}(\boldsymbol{y})
+R^{(1)}_{\mathrm{inv}(1)}(\boldsymbol{y})
+R^{(1)}_{\mathrm{inv}(2)}(\boldsymbol{y})\Big)
+\mathcal{O}(\epsilon^2),
\end{split}
\end{equation}
\end{widetext}
where each coefficient is expressed in terms of the coefficients of $R$ as
\begin{widetext}
\begin{equation}
\label{inv2_coeff_all_in_one}
\begin{split}
R^{(-1)}_{\mathrm{inv}(0)}
&=
\Big(R^{(1)}_{(0)}\Big)^{-1},
\\
R^{(0)}_{\mathrm{inv}(0)}(\boldsymbol{y})
&=
-\Big(R^{(1)}_{(0)}\Big)^{-1}\,
R^{(2)}_{(0)}(\boldsymbol{y})\,
\Big(R^{(1)}_{(0)}\Big)^{-1},
\\
R^{(0)}_{\mathrm{inv}(1)}(\boldsymbol{y})
&=
-\Big(R^{(1)}_{(0)}\Big)^{-1}\,
R^{(2)}_{(1)}(\boldsymbol{y})\,
\Big(R^{(1)}_{(0)}\Big)^{-1},
\\
R^{(1)}_{\mathrm{inv}(0)}(\boldsymbol{y})
&=
\Big(R^{(1)}_{(0)}\Big)^{-1}
R^{(2)}_{(0)}(\boldsymbol{y})
\Big(R^{(1)}_{(0)}\Big)^{-1}
R^{(2)}_{(0)}(\boldsymbol{y})
\Big(R^{(1)}_{(0)}\Big)^{-1}
-
\Big(R^{(1)}_{(0)}\Big)^{-1}
R^{(3)}_{(0)}(\boldsymbol{y})
\Big(R^{(1)}_{(0)}\Big)^{-1},
\\
R^{(1)}_{\mathrm{inv}(1)}(\boldsymbol{y})
&=
\Big(R^{(1)}_{(0)}\Big)^{-1}
R^{(2)}_{(0)}(\boldsymbol{y})
\Big(R^{(1)}_{(0)}\Big)^{-1}
R^{(2)}_{(1)}(\boldsymbol{y})
\Big(R^{(1)}_{(0)}\Big)^{-1}
+
\Big(R^{(1)}_{(0)}\Big)^{-1}
R^{(2)}_{(1)}(\boldsymbol{y})
\Big(R^{(1)}_{(0)}\Big)^{-1}
R^{(2)}_{(0)}(\boldsymbol{y})
\Big(R^{(1)}_{(0)}\Big)^{-1}
\\&+
\Big(R^{(1)}_{(0)}\Big)^{-1}\,
R^{(3)}_{(1)}(\boldsymbol{y})\,
\Big(R^{(1)}_{(0)}\Big)^{-1},
\\
R^{(1)}_{\mathrm{inv}(2)}(\boldsymbol{y})
&=
\Big(R^{(1)}_{(0)}\Big)^{-1}
R^{(2)}_{(1)}(\boldsymbol{y})
\Big(R^{(1)}_{(0)}\Big)^{-1}
R^{(2)}_{(1)}(\boldsymbol{y})
\Big(R^{(1)}_{(0)}\Big)^{-1}
-
\Big(R^{(1)}_{(0)}\Big)^{-1}
R^{(3)}_{(2)}(\boldsymbol{y})
\Big(R^{(1)}_{(0)}\Big)^{-1}.
\end{split}
\end{equation}
\end{widetext}

We next expand the determinant of the Jacobian matrix appearing in the learning-rate formula.
Using
\begin{equation}
\frac{\partial}{\partial y_l}\ln\big|\det R\big|
=
\mathrm{tr}\!\left(R^{-1}\frac{\partial R}{\partial y_l}\right),
\end{equation}
and the expansions Eqs. (\ref{asymptotic_jacobian}) and (\ref{asymptotic_jacobian_inv}),
we obtain an $\epsilon$-expansion of $\partial_{y_l}\ln|\det R|$ as
\begin{equation}
\frac{\partial}{\partial y_l}\ln\big|\det R\big|
=
\epsilon\,\Gamma^{(1)}_{(0)l}
+\epsilon^2\left(\Gamma^{(2)}_{(0)l}+\Gamma^{(2)}_{(1)l}\right)
+\mathcal{O}(\epsilon^3),
\end{equation}
where
\begin{widetext}
\begin{equation}
\label{asymptotic_logdetR_expand}
\begin{split}
\Gamma^{(1)}_{(0)l}
&=
\mathrm{tr}\!\left[
R^{(-1)}_{\mathrm{inv}(0)}
\left(
\frac{\partial R^{(2)}_{(1)}(\boldsymbol{y})}{\partial y_l}
\right)
\right],
\\
\Gamma^{(2)}_{(0)l}
&=
\mathrm{tr}\!\left[
R^{(-1)}_{\mathrm{inv}(0)}
\left(
\frac{\partial R^{(3)}_{(1)}(\boldsymbol{y})}{\partial y_l}
\right)
\right]
+\mathrm{tr}\!\left[
\Big(R^{(0)}_{\mathrm{inv}(0)}(\boldsymbol{y})\Big)
\left(\frac{\partial R^{(2)}_{(1)}(\boldsymbol{y})}{\partial y_l}
\right)
\right]
\\
\Gamma^{(2)}_{(1)l}
&=
\mathrm{tr}\!\left[
R^{(-1)}_{\mathrm{inv}(0)}
\left(
\frac{\partial R^{(3)}_{(2)}(\boldsymbol{y})}{\partial y_l}
\right)
\right]
+\mathrm{tr}\!\left[
\Big(R^{(0)}_{\mathrm{inv}(1)}(\boldsymbol{y})\Big)
\left(\frac{\partial R^{(2)}_{(1)}(\boldsymbol{y})}{\partial y_l}
\right)
\right].
\end{split}
\end{equation}
\end{widetext}
By substituting Eqs. (\ref{asymptotic_stationary}) -- (\ref{asymptotic_jacobian_inv}) into \eqref{transformation_learning_asymptotic}
and collecting terms order by order in $\epsilon$,
we obtain the following expansions for $l^i_{p}$ and $l^i_{R}$:
\begin{widetext}
\begin{equation}
\label{lp_expand_def}
l^i_{p}
=
\epsilon^4 V \int d \boldsymbol{y}\,
\sum_{k,l,m}
\left(A^i_{klm}y_m+\epsilon B^i_{klm}y_m+\epsilon^2 C^i_{klm}y_m+\epsilon^2\sum_{n,o}D^i_{klmno}y_m y_n y_o +\mathcal{O}(\epsilon^3)\right)
y_l \left(2\chi \alpha +4\beta|\boldsymbol{y}|^2\right)\tilde{p}_{\mathrm{st}}(\boldsymbol{y}),
\end{equation}
\end{widetext}
where the coefficients are defined by
\begin{widetext}
\begin{equation}
\label{ABCD_def}
\begin{split}
\sum_{m}A^i_{klm}y_m
&=
R^{(1)}_{(0)ik}\,
\tilde{\nu}^{(0)}_{(1)k}\,
R^{(-1)}_{\mathrm{inv}(0)li},\\
\sum_{m}B^i_{klm}y_m
&=
R^{(2)}_{(0)ik}\,
\tilde{\nu}^{(0)}_{(1)k}\,
R^{(-1)}_{\mathrm{inv}(0)li}
+
R^{(1)}_{(0)ik}\,
\tilde{\nu}^{(0)}_{(1)k}\,
R^{(0)}_{\mathrm{inv}(0)li},\\
\sum_{m}C^i_{klm}y_m
&=
\tilde{\nu}^{(0)}_{(1)k}\left(
R^{(1)}_{(0)ik}
R^{(1)}_{\mathrm{inv}(0)li}
+
R^{(2)}_{(0)ik}
R^{(0)}_{\mathrm{inv}(0)li}
+
R^{(3)}_{(0)ik}
R^{(-1)}_{\mathrm{inv}(0)li}
\right)
+
R^{(1)}_{(0)ik}\,
\tilde{\nu}^{(2)}_{(1)k}\,
R^{(-1)}_{\mathrm{inv}(0)li},\\
\sum_{m,n,o}D^i_{klmno}y_m y_n y_o
&=
\tilde{\nu}^{(0)}_{(1)k}\left(
R^{(1)}_{(0)ik}
R^{(1)}_{\mathrm{inv}(2)li}
+
R^{(2)}_{(1)ik}
R^{(0)}_{\mathrm{inv}(1)li}
+
R^{(3)}_{(2)ik}
R^{(-1)}_{\mathrm{inv}(0)li}
\right)
+
R^{(1)}_{(0)ik}\,
\tilde{\nu}^{(2)}_{(3)k}\,
R^{(-1)}_{\mathrm{inv}(0)li}.
\end{split}
\end{equation}
\end{widetext}

Similarly, the contribution originating from the determinant of the Jacobian matrix can be written as
\begin{equation}
\label{lR_expand_def}
l^i_{R}
=
-\epsilon^2 \int d \boldsymbol{y}\,
\left(\sum_{k,l,m,n} E^i_{klmn}y_m y_n +\mathcal{O}(\epsilon)\right)\tilde{p}_{\mathrm{st}}(\boldsymbol{y}),
\end{equation}
where
\begin{widetext}
\begin{equation}
\label{E_def}
\sum_{m,n} E^i_{klmn}y_m y_n
=
\Gamma^{(1)}_{(0)l}
\left(
R^{(1)}_{(0)ik}
\tilde{\nu}^{(0)}_{(1)k}
R^{(0)}_{\mathrm{inv}(1)li}
+
R^{(2)}_{(1)ik}
\tilde{\nu}^{(0)}_{(1)k}
R^{(-1)}_{\mathrm{inv}(0)li}
\right)
+
\Gamma^{(2)}_{(1)l}\,
R^{(1)}_{(0)ik}
\tilde{\nu}^{(0)}_{(1)k}
R^{(-1)}_{\mathrm{inv}(0)li}.
\end{equation}
\end{widetext}

We next evaluate the Gaussian averages appearing in $l^i_p$ and $l^i_R$ in the stationary regime and in the oscillatory regime.
We first consider the stationary regime ($\chi=-1$), where the distribution is concentrated around the origin.
In this regime, the Cartesian moments are given by
\begin{equation}
\label{moments_cartesian_chi_minus_only}
\begin{split}
\langle -2\alpha y_m y_l+4\beta y_m y_l|\boldsymbol y|^2 \rangle
&=
-\frac{\delta_{ml}}{\epsilon^4 V }
+
 \mathcal{O}(V^{-2}).
\end{split}
\end{equation}
Terms such as $\langle y_m y_n y_o y_l\rangle$ and $\langle y_m y_n y_o y_l|\boldsymbol y|^2\rangle$ contribute starting from $\mathcal{O}(V^{-2})$ and therefore do not affect the leading-order terms.

By inserting these moments into \eqref{lp_expand_def}, we obtain the asymptotic expansions
\begin{equation}
\begin{split}
l^i_p
&=
-\sum_{k,l}A^i_{kll}
-\epsilon \sum_{k,l}B^i_{kll}
-\epsilon^2 \sum_{k,l}C^i_{kll}+\mathcal{O}(\epsilon^{3}),
\end{split}
\end{equation}
and $l^i_R$ contributes $\mathcal{O}(V^{-1})$ and therefore does not affect the leading-order terms.

We next evaluate the Gaussian averages in the oscillatory regime ($\chi=1$). In this regime, the Cartesian moments are given by
\begin{widetext}
\begin{equation}
\label{moments_cartesian_chi_plus_only}
\begin{split}
\langle 2\alpha y_m y_l+4\beta y_m y_l|\boldsymbol y|^2 \rangle
&=
-\frac{\delta_{ml}}{\epsilon^4 V},\\
\langle 2\alpha y_m y_n y_o y_l +4 \beta y_m y_n y_o y_l|\boldsymbol y|^2\rangle
&=
\frac{\alpha}{4\beta \epsilon^4 V }
\Big(
\delta_{mn}\delta_{ol}
+\delta_{mo}\delta_{nl}
+\delta_{ml}\delta_{no}
\Big).
\end{split}
\end{equation}
\end{widetext}
Substituting these moments into \eqref{lp_expand_def} and \eqref{lR_expand_def}, we obtain
\begin{widetext}
\begin{equation}
\begin{split}
l^i_p
&=
-\sum_{k,l}A^i_{kll}
-\epsilon \sum_{k,l}B^i_{kll}
-\epsilon^2 \sum_{k,l}C^i_{kll}
+\epsilon^2\frac{\alpha}{4\beta}\sum_{k,l,m}\left(D^i_{klmml}+D^i_{klmlm}+D^i_{kllmm}\right)
+\mathcal{O}(\epsilon^{3}),
\end{split}
\end{equation}
\end{widetext}
and
\begin{equation}
\begin{split}
l^i_R
&=
\epsilon^2\frac{\alpha}{4\beta}
\sum_{k,l,m}E^i_{klmm}
+\mathcal{O}(\epsilon^{3}).
\end{split}
\end{equation}
Therefore, the learning rate in the deterministic limit $V\rightarrow \infty$ is expressed as 
\begin{widetext}
\begin{equation}
\label{lst_general}
\begin{split}
\lim_{V\rightarrow \infty}
l^i_{\mathrm{st}}(\epsilon,V)
=
\begin{cases}
-\displaystyle\sum_{k,l}A^i_{kll}
-\epsilon \displaystyle\sum_{k,l}B^i_{kll}
-\epsilon^2 \displaystyle\sum_{k,l}C^i_{kll}
+\mathcal{O}(\epsilon^3)
& \mathrm{for} \quad \chi=-1,\\[1mm]
-\displaystyle\sum_{k,l}A^i_{kll}
-\epsilon \displaystyle\sum_{k,l}B^i_{kll}
-\epsilon^2 \displaystyle\sum_{k,l}C^i_{kll}
+\epsilon^2 \dfrac{\alpha}{4\beta}
\displaystyle\sum_{k,l,m}\!\left(D^i_{klmml}+D^i_{klmlm}+D^i_{kllmm}\right)\\
\qquad
+\epsilon^2\dfrac{\alpha}{4\beta}\displaystyle\sum_{k,l,m}E^i_{klmm}
+\mathcal{O}(\epsilon^3)
& \mathrm{for} \quad \chi=1,
\end{cases}
\end{split}
\end{equation}
\end{widetext}
where $\alpha$, $\beta$, $A^i_{kll}$,  $B^i_{kll}$, $C^i_{kll}$, $D^i_{klmlm}$ and $E^i_{klmm}$ are constants independent of the variables $\boldsymbol{y}$, the small parameter $\epsilon$ and the system size $V$. 
In the stationary regime $(\chi=-1)$, the deterministic limit is governed by the leading-order contribution obtained from the linearized dynamics, with corrections of order $\mathcal{O}(\epsilon^{3})$.
In contrast, in the oscillatory regime $(\chi=1)$ additional $\epsilon^{2}$ terms appear and remain finite as $V\to\infty$. This results in a non-smooth change in the learning rate at the bifurcation point in the deterministic limit.
In particular, for the Brusselator with $a = 1$, \eqref{lst_general} yields \eqref{lst_piecewise}.

\section{Parameters of the Brusselator}\noindent
\label{Brusselator}
In this section, we give the explicit forms of the constants and vectors resulting from the singular perturbation analysis for the Brusselator.
The bifurcation parameter is defined as
\begin{equation}
\mu
\equiv
\frac{b - b_c}{b_c}.
\end{equation}
From the linear stability analysis, we obtain
\begin{equation}
    L_0
    =
    \left(
    \begin{matrix}
        a^2&a^2\\
        -(a^2+1)&-a^2
    \end{matrix}
    \right),\quad
    \lambda_0 = \mathrm{i}\omega_0=\mathrm{i}a
\end{equation}
and the corresponding right and left eigenvectors as
\begin{equation}
    \vec{U}
    =
    \left(
    \begin{matrix}
        1\\-1+\mathrm{i}\frac{1}{a}
    \end{matrix}
    \right),\quad
    \vec{U}^{\ast}
    =
    \frac{1}{2}
    \left(
    1-\mathrm{i}a,-\mathrm{i}a
    \right).
\end{equation}
Then, from the linear term of order $\epsilon^2$ and \eqref{lambda_1}, we obtain
\begin{equation}
    L_1
    =
    (a^2+1)
    \left(
    \begin{matrix}
        1&0\\
        -1&0
    \end{matrix}
    \right),\quad
    \sigma_1
    =
    \frac{a^2+1}{2}
    ,\quad
    \omega_1=0.
\end{equation}
From the time evolution equations of the concentrations, for any vectors $\vec{\alpha}, \vec{\beta}, \vec{\gamma}$, the operators $M_0$, $N_0$ and $M_1$ are calculated as
\begin{equation}
\begin{split}
    (M_0 : \vec{\alpha}\vec{\beta})_1
    &=
    \frac{a^2+1}{a^2}\alpha_1\beta_1+a(\alpha_1\beta_2+\alpha_2\beta_1)\\&=-(M_0\vec{\alpha}\vec{\beta})_2\\
\end{split}
\end{equation}
\begin{equation}
\begin{split}(N_0 : \vec{\alpha}\vec{\beta}\vec{\gamma})_1
    &=
    \frac{1}{3}(\alpha_1\beta_1\gamma_2+\alpha_1\beta_2\gamma_1+\alpha_2\beta_1\gamma_1),\\
    &=
    -(N_0\vec{\alpha}\vec{\beta}\vec{\gamma})_2
\end{split}
\end{equation}
\begin{equation}
\begin{split}
    (M_1 : \vec{\alpha}\vec{\beta})_1,
    &=
    \frac{a^2+1}{a^2}\alpha_1\beta_1=-(M_1 : \vec{\alpha}\vec{\beta})_2.
\end{split}
\end{equation}
The constant vectors $\vec{V}_{+}$, $\vec{V}_{-}$ and $\vec{V}_0$ defined in \eqref{V_plus} and \eqref{V_0} are given by
\begin{equation}
\begin{split}
    &\vec{V}_+
    =
    \vec{V}_-
    =
    \frac{(1+\mathrm{i}a)^2}{3a^3}
    \left(
    \begin{matrix}
        -2\mathrm{i}a\\
        1+2\mathrm{i}a
    \end{matrix}
    \right),\\
    &\vec{V}_0
    =
    \frac{2(a^2-1)}{a^3}
    \left(
    \begin{matrix}
        0\\1
    \end{matrix}
    \right).
\end{split}
\end{equation}
Using the above terms, we obtain \eqref{third_g} as
\begin{equation}
\begin{split}
    g
    &=
    g_1+\mathrm{i} g_2\\
    &=
    \frac{1}{2}
    \left(
    \frac{a^2+2}{a^2}+\mathrm{i}\frac{4a^4-7a^2+4}{3a^3}
    \right).
\end{split}
\end{equation}
In computing the third-order term of the variable transformation, we employ the pseudo-inverse matrix:
\begin{equation}
    G
    =
    \frac{1}{(2a^2+1)^2}
    \left(
    \begin{matrix}
        -\mathrm{i}a-a^2&a^2+1\\
        -a^2&-\mathrm{i}a+a^2
    \end{matrix}
    \right).
\end{equation}
Furthermore, the diffusion matrix in the Fokker–Planck equation \eqref{normal_form_diffusion} can be calculated using the previously introduced quantities, yielding
\begin{equation}
    \tilde{D}
    =
    \frac{a}{2}
    \left(
    \begin{matrix}
        a^2+2&-a\\
        -a&a^2
    \end{matrix}
    \right).
\end{equation}
\\

\begin{thebibliography}{99}

\bibitem{Sekimoto2010} K. Sekimoto, Concept of Heat on Mesoscopic Scales, in \textit{Stochastic Energetics} (Springer, Berlin, Heidelberg, 2010), pp.\ 135--174.

\bibitem{parrondo2015thermodynamics} J. M. R. Parrondo, J. M. Horowitz, and T. Sagawa, Thermodynamics of information, Nat. Phys. \textbf{11}, 131--139 (2015).

\bibitem{Sagawa2019} T. Sagawa, Second Law, Entropy Production, and Reversibility in Thermodynamics of Information, in \textit{Energy Limits in Computation: A Review of Landauer's Principle, Theory and Experiments}, eds.\ C. S. Lent, A. O. Orlov, W. Porod, and G. L. Snider (Springer, Cham, 2019), pp.\ 101--139.

\bibitem{Shannon} C. E. Shannon, A mathematical theory of communication, Bell Syst. Tech. J. \textbf{27}, 379--423 (1948).

\bibitem{coverthomas} T. M. Cover and J. A. Thomas, \textit{Elements of Information Theory} (John Wiley \& Sons, 1999).

\bibitem{szilard} L. Szilard, On the reduction of entropy in a thermodynamic system by the interference of an intelligent being, Z. Phys. \textbf{53}, 840--856 (1929).

\bibitem{Landauer} R. Landauer, Irreversibility and Heat Generation in the Computing Process, IBM J. Res. Dev. \textbf{5}, 183--191 (1961).

\bibitem{Sagawa} T. Sagawa and M. Ueda, Nonequilibrium thermodynamics of feedback control, Phys. Rev. E \textbf{85}, 021104 (2012).

\bibitem{sagawa_ueda_2010} T. Sagawa and M. Ueda, Generalized Jarzynski Equality under Nonequilibrium Feedback Control, Phys. Rev. Lett. \textbf{104}, 090602 (2010).

\bibitem{TUR} A. C. Barato and U. Seifert, Thermodynamic Uncertainty Relation for Biomolecular Processes, Phys. Rev. Lett. \textbf{114}, 158101 (2015).

\bibitem{Tanogami_Saito_2023} T. Tanogami, T. V. Vu, and K. Saito, Universal bounds on the performance of information-thermodynamic engine, Phys. Rev. Res. \textbf{5}, 043280 (2023).

\bibitem{langevin_information} A. E. Allahverdyan, D. Janzing, and G. Mahler, Thermodynamic efficiency of information and heat flow, J. Stat. Mech.: Theory Exp. \textbf{2009}, P09011 (2009).

\bibitem{conti_information} J. M. Horowitz and M. Esposito, Thermodynamics with Continuous Information Flow, Phys. Rev. X \textbf{4}, 031015 (2014).

\bibitem{Multipartite_information_flow} J. M. Horowitz, Multipartite information flow for multiple Maxwell demons, J. Stat. Mech.: Theory Exp. \textbf{2015}, P03006 (2015).

\bibitem{Barato_2014} A. C. Barato, D. Hartich, and U. Seifert, Efficiency of cellular information processing, New J. Phys. \textbf{16}, 103024 (2014).

\bibitem{Hartich_2014} D. Hartich, A. C. Barato, and U. Seifert, Stochastic thermodynamics of bipartite systems: transfer entropy inequalities and a Maxwell’s demon interpretation, J. Stat. Mech.: Theory Exp. \textbf{2014}, P02016 (2014).

\bibitem{ito2015maxwell} S. Ito and T. Sagawa, Maxwell’s demon in biochemical signal transduction with feedback loop, Nat. Commun. \textbf{6}, 7498 (2015).

\bibitem{Sensory_capacity} D. Hartich, A. C. Barato, and U. Seifert, Sensory capacity: An information theoretical measure of the performance of a sensor, Phys. Rev. E \textbf{93}, 022116 (2016).

\bibitem{Role} T. Matsumoto and T. Sagawa, Role of sufficient statistics in stochastic thermodynamics and its implication to sensory adaptation, Phys. Rev. E \textbf{97}, 042103 (2018).

\bibitem{Information_Arbitrage} M. P. Leighton, J. Ehrich, and D. A. Sivak, Information Arbitrage in Bipartite Heat Engines, Phys. Rev. X \textbf{14}, 041038 (2024).

\bibitem{Prigogine_Nicolis_1971} I. Prigogine and G. Nicolis, Biological order, structure and instabilities, Q. Rev. Biophys. \textbf{4}, 107--148 (1971).

\bibitem{novak2008design} B. Nov\'{a}k and J. J. Tyson, Design principles of biochemical oscillators, Nat. Rev. Mol. Cell Biol. \textbf{9}, 981--991 (2008).

\bibitem{cell_rythms} A. Goldbeter, \textit{Biochemical Oscillations and Cellular Rhythms: The Molecular Bases of Periodic and Chaotic Behaviour} (Cambridge University Press, Cambridge, 1996).

\bibitem{Circadian_Oscillation} M. Nakajima, K. Imai, H. Ito, T. Nishiwaki, Y. Murayama, H. Iwasaki, T. Oyama, and T. Kondo, Reconstitution of Circadian Oscillation of Cyanobacterial KaiC Phosphorylation in Vitro, Science \textbf{308}, 414--415 (2005).

\bibitem{cell_cycle} J. E. Ferrell Jr., T. Yu-Chen Tsai, and Q. Yang, Modeling the Cell Cycle: Why Do Certain Circuits Oscillate?, Cell \textbf{144}, 874--885 (2011).

\bibitem{phase_transition} B. Nguyen, U. Seifert, and A. C. Barato, Phase transition in thermodynamically consistent biochemical oscillators, J. Chem. Phys. \textbf{149}, 045101 (2018).

\bibitem{Coherence_of_biochemical} A. C. Barato and U. Seifert, Coherence of biochemical oscillations is bounded by driving force and network topology, Phys. Rev. E \textbf{95}, 062409 (2017).

\bibitem{Design_priciples_for_biochemical} Z. Cao, H. Jiang, and Z. Hou, Design principles for biochemical oscillations with limited energy resources, Phys. Rev. Res. \textbf{2}, 043331 (2020).

\bibitem{Robust_oscillations} C. del Junco and S. Vaikuntanathan, Robust oscillations in multi-cyclic Markov state models of biochemical clocks, J. Chem. Phys. \textbf{152}, 055101 (2020).

\bibitem{Stochastic_Thermodynamics_in_Mesoscopic} T. Xiao, Z. Hou, and H. Xin, Stochastic Thermodynamics in Mesoscopic Chemical Oscillation Systems, J. Phys. Chem. B \textbf{113}, 9316--9320 (2009).

\bibitem{The_free-energy_cost} Y. Cao, H. Wang, Q. Ouyang, and Y. Tu, The free-energy cost of accurate biochemical oscillations, Nat. Phys. \textbf{11}, 772--778 (2015).

\bibitem{The_energy_cost_and_optimal_design} D. Zhang, Y. Cao, Q. Ouyang, and Y. Tu, The energy cost and optimal design for synchronization of coupled molecular oscillators, Nat. Phys. \textbf{16}, 95--100 (2020).

\bibitem{Improved_estimation_for_energy} Z. Cao and Z. Hou, Improved estimation for energy dissipation in biochemical oscillations, J. Chem. Phys. \textbf{157}, 025102 (2022).

\bibitem{Strogatz} S. H. Strogatz, \textit{Nonlinear Dynamics and Chaos: With Applications to Physics, Biology, Chemistry, and Engineering} (Westview Press, 2000).

\bibitem{Kuramoto1984} Y. Kuramoto, \textit{Chemical Oscillations, Waves, and Turbulence} (Springer, Berlin, Heidelberg, 1984).

\bibitem{Baras1982}
F.~Baras, M.~Malek~Mansour, and C.~Van~den~Broeck,
``Asymptotic properties of coupled nonlinear Langevin equations in the limit of weak noise. II: Transition to a limit cycle,''
\textit{Journal of Statistical Physics} \textbf{28}(3), 577--587 (1982).
doi:10.1007/BF01008325

\bibitem{Knobloch1983}
E.~Knobloch and K.~A.~Wiesenfeld,
``Bifurcations in fluctuating systems: The center-manifold approach,''
\textit{Journal of Statistical Physics} \textbf{33}(3), 611--637 (1983).
doi:10.1007/BF01018837

\bibitem{Arnold1996}
L.~Arnold, N.~Sri Namachchivaya, and K.~R.~Schenk-Hopp\'{e},
``Toward an Understanding of Stochastic Hopf Bifurcation,''
\textit{International Journal of Bifurcation and Chaos} \textbf{6}(11), 1947--1975 (1996).
doi:10.1142/S0218127496001272


\bibitem{Gillespie_langevin} D. T. Gillespie, The chemical Langevin equation, J. Chem. Phys. \textbf{113}, 297--306 (2000).

\bibitem{gardiner1985handbook} C. W. Gardiner, \textit{Handbook of Stochastic Methods} (Springer, Berlin, 1985).
\bibitem{non-bipartite} R. Ch\'{e}trite, M. L. Rosinberg, T. Sagawa, and G. Tarjus, Information thermodynamics for interacting stochastic systems without bipartite structure, J. Stat. Mech.: Theory Exp. \textbf{2019}, 114002 (2019).

\bibitem{Freitas2023}
N.~Freitas and M.~Esposito,
``Information flows in macroscopic Maxwell's demons,''
\textit{Phys. Rev. E} \textbf{107}, 014136 (2023).
\bibitem{GithubData}
K. Matsumoto, Processed datasets and Julia scripts for ``Singularity of information flow at the Hopf bifurcation point,'' GitHub repository, \url{https://github.com/Ken-Matsumoto-code/hopf-learning-rate}.
\bibitem{irriversible_circulation} K. Tomita and H. Tomita, Irreversible Circulation of Fluctuation, Prog. Theor. Phys. \textbf{51}, 1731--1749 (1974).


\end{thebibliography}

\end{document}